\documentclass[pra,floatfix,twocolumn,amsfonts,amsmath,amssymb,nofootinbib,reprint,superscriptaddress,showkeys]{revtex4-2}

\usepackage{adjustbox}

\usepackage[T1]{fontenc}
\usepackage[latin1]{inputenc}
\usepackage{amsmath,graphicx}
\usepackage{amssymb,bm}
\usepackage{epsfig,graphicx}
\usepackage{mathrsfs}
\usepackage[active]{srcltx}
\usepackage{mdframed}
\usepackage{mathtools}
\usepackage{xcolor}
\usepackage{hyperref}
\usepackage{braket}
\usepackage{gensymb}
\usepackage{enumerate}
\usepackage{epsfig,libertine,dsfont,amsthm}
\usepackage{layouts}
\usepackage[justification=Justified]{subcaption}
\newtheorem{cor}{Corollary}
\newtheorem{lem}{Lemma}
\newtheorem{prop}{Proposition}

\newtheorem{theo}{Theorem}

\newtheorem{defin}{Definition}

\newtheorem{rem}{Remark}

\usepackage{hyperref}
\hypersetup{colorlinks=true,urlcolor=red,citecolor=green,filecolor=magenta}

\begin{document}

\title{Universal Quantum Computation via Superposed Orders of Single-Qubit Gates}

\author{Kyrylo Simonov}
\email{kyrylo.simonov@univie.ac.at}
\affiliation{%
  %\institution{
  Fakult\"at f\"ur Mathematik, Universit\"at Wien, Oskar-Morgenstern-Platz 1, %}
 % \city{
 1090 Vienna, %}
  %\country{
  Austria}%}

\author{Marcello Caleffi}
\email{marcello.caleffi@unina.it}
\affiliation{
%\institution{
\href{www.quantuminternet.it}{www.QuantumInternet.it} Research Group, University of Naples Federico II, 
%}
%\city{
Naples, %}
%\country{
Italy}%}

\author{Jessica Illiano}
\email{jessica.illiano@unina.it}
\affiliation{
%\institution{
\href{www.quantuminternet.it}{www.QuantumInternet.it} Research Group, University of Naples Federico II, 
%}
%\city{
Naples, %}
%\country{
Italy}%}

\author{Jacquiline Romero}
\email{m.romero@uq.edu.au}
\affiliation{
%\institution{
Australian Research Council Centre of Excellence for Engineered Quantum Systems \& School of Mathematics and Physics, University of Queensland, %}
%\city{
St. Lucia, Brisbane QLD 4072, %}
%\country{
Australia}%}

\author{Angela Sara Cacciapuoti}
\email{angelasara.cacciapuoti@unina.it}
\affiliation{
%\institution{
\href{www.quantuminternet.it}{www.QuantumInternet.it} Research Group, University of Naples Federico II, 
%}
%\city{
Naples, %}
%\country{
Italy}%}

%\renewcommand{\shortauthors}{Authors et al.}
\begin{abstract}
  Superposed orders of quantum channels have already been proved -- both theoretically and experimentally -- to enable unparalleled opportunities in the quantum communication domain. Superposition of orders can be exploited within the quantum computing domain as well, by relaxing the (traditional) assumption underlying quantum computation about applying gates in a well-defined causal order. In this context, we address a fundamental question arising with quantum computing: whether superposed orders of single-qubit gates can enable universal quantum computation. As shown in this paper, the answer to this key question is a definitive ``\textit{yes}''. Indeed, we prove that any two-qubit controlled quantum gate can be \textit{deterministically} realized, including the so-called Barenco gate that alone enables universal quantum computation.
\end{abstract}

%    <ccs2012>
 %       <concept>
            %<concept_id>10003752.10003753.10003758</concept_id>
 %           <concept_desc>Theory of computation~Quantum computation theory</concept_desc>
            %<concept_significance>500</concept_significance>
  %      </concept>
            %<concept_id>10010520.10010521.10010542.10010550</concept_id>
 %           <concept_desc>Computer systems organization~Quantum computing</concept_desc>
            %<concept_significance>500</concept_significance>
 %       </concept>
 %   </ccs2012>
%\end{CCSXML}

%\ccsdesc[500]{Theory of computation~Quantum computation theory}
%\ccsdesc[500]{Computer systems organization~Quantum computing}

\keywords{Quantum computing architecture, Fault-tolerant Quantum Computation, Universal Quantum Computing, Quantum Computation, Quantum Gates, Quantum Switch, Superposed Orders, ERC-CoG grant QNattyNet}

%\received{01 January 2024}

\maketitle

\section{Introduction}
\label{sec:1}

\begin{figure*}
\centering
%\begin{tabular}{ cc }
    \begin{minipage}{0.45\linewidth}
    	\centering
    	\begin{adjustbox}{width=0.6\linewidth} 
 		\centering
		\includegraphics[width=\columnwidth]{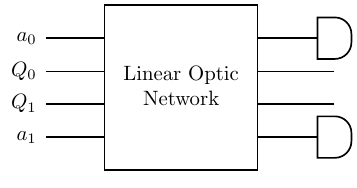}
        \end{adjustbox}
    \subcaption{Simplified diagram of a linear optic implementation of a $\mathtt{CNOT}$ logic gate, where $Q_0$ and $Q_1$ represent the control qubit and the target qubit and $a_0$, $a_1$ are two ancillary photons.}
    \label{Fig:02a}
    \end{minipage}
    %&
    \begin{minipage}{0.05\linewidth}
    \begin{tabular}{cc}
    &
    \end{tabular}
    \end{minipage}	
    \begin{minipage}{0.45\linewidth}
		\centering
		\includegraphics[width=\linewidth]{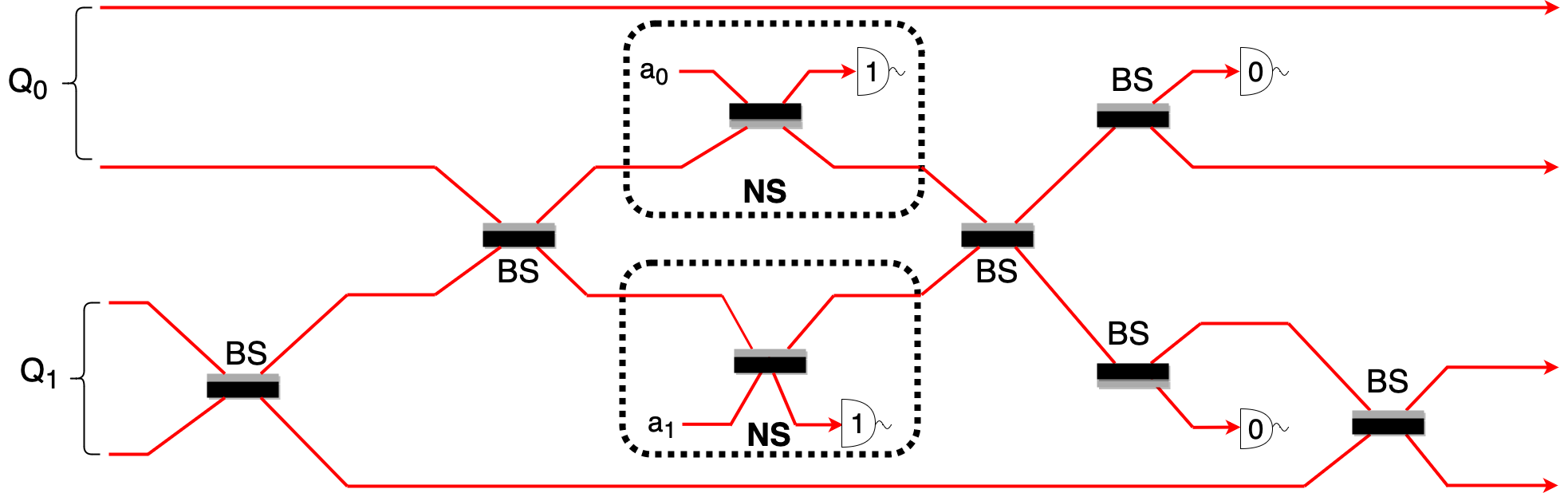}
    \subcaption{KLM $\mathtt{CNOT}$ scheme with simplified NS gates. $Q_0$ and $Q_1$ denote the control qubit and the target qubit, respectively. This scheme assumes the logical qubits to be encoded through spatial modes (path encoding).}
    \label{Fig:02b}
	\end{minipage}\\
%	\end{tabular}
    \caption{Non-deterministic $\mathtt{CNOT}$ gate via linear optics.}
    %\Description{Non-deterministic $\mathtt{CNOT}$ gate via linear optics.}
    \label{Fig:02}
    %\hrulefill
\end{figure*}

During the last ten years, there has been an increasing widespread interest on investigating the advantages arising from the superposition of traversing orders of quantum channels in the quantum communication domain, due to the outstanding possibilities arising from the non-classical propagation of quantum carriers \cite{CalCac-25, ChiKri-19, AbbWecHor-20, KriChiSal-20, Chiribella2021,CacPelIll-25,PelCalCac-25}. While classical carriers propagate through classical communication paths, quantum channels can be placed in a genuinely quantum setting with no counterpart in classical world \cite{Chiribella2012, OreCosBru-12}, giving rise to the concept of \textit{quantum path} \cite{IllCacMan-22,CacIllKou-22,CacIllCal-23}. This non-classical propagation has been experimentally validated \cite{ProMoqAra-15,RubRozFei-17}, and it finds a possible physical realization in the device known as \textit{quantum switch} \cite{GosGiaKew-18,Guo2020}, which enables a coherent superposition of alternative channel configurations. Specifically, by the means of the quantum switch, the quantum carrier can travel through different communication channels in a quantum superposition of different causal orders. This, in turn, makes the order of the communication channels indefinite and it returns disruptive advantages, for instance, for communications through noisy channels \cite{CalCac-20,ChaCalCac-21,CalSimCac-23}. Furthermore, it has been recently demonstrated that the quantum switch can be exploited for generating different classes of genuine multipartite entangled states starting from separable inputs \cite{Koudia2021,Chen2021}. Ongoing research is exploring further intriguing aspects of the unconventional resources enabled by the quantum switch \cite{Felce2020, Guha2020, ChiWilCha-21, Simonov2022_Erg, Milz2022, maity2024activating, mukherjee2024interplay, Simonov2022}. Introducing the quantum switch to the engineering of quantum communications present key advantages.

The possibilities enabled by the superposition of orders of quantum operations can be exploited within the quantum computing domain as well \cite{ChiDarPerVal-09, ChiDarPer-13}. Specifically, it is possible to relax the (traditional) assumption underlying quantum computation, i.e., quantum gates applied in a well-defined causal order only \cite{Wechs2021}. By doing this, a novel and more-general computing framework arises, which has been applied to a number of problems, both theoretically \cite{ChiDarPerVal-09,ColDarFac-12,ChiDarPer-13,Araujo2014,TadNerAol-19,RenBru-22,EscDelWal-23,LiuMenSon-23} and experimentally \cite{ProMoqAra-15, Taddei2021}. 

In this context, we address a fundamental question arising with quantum computing, namely, whether superposed orders of quantum operations can constitute a novel paradigm for \textit{universal quantum computing}.

As proved in the following, the answer to this key question is a definitive ``yes''. Specifically, through the manuscript we show that any controlled quantum gate can be \textit{deterministically} realized via superposition of the traversing orders of single-qubit gates. Notably, this includes the two-qubit gate known as \textit{Barenco gate}, which \textit{alone} enables universal quantum computation \cite{Barenco1995}.

% --------------------------------------------------------------------------
\subsection{Contribution}
\label{sec:1.1}
Our contributions can be summarized as follows:
\begin{itemize}
    \item we provide a general framework for the realization of controlled quantum gates through superposed orders of single-qubit unitaries;
    \item we prove that this framework enables the \textit{deterministic} realization of any two-qubit controlled gate;
    \item we specialize the framework for two-qubit controlled gates widely used in quantum computing, including $\mathtt{CNOT}$, $\mathtt{CZ}$, and the \textit{universal} Barenco gate. Furthermore, we extend the implementation of controlled gates to $d$-dimensional systems.
\end{itemize}

We note that the relevance of our work can be linked to the photonic computing domain considering physical implementations of the quantum switch are mostly photonic. As a matter of fact, photonic quantum gates are either probabilistic or based on pre-shared multipartite-entangled states, as discussed in Section~\ref{sec:2.1}. Conversely, our framework enables the deterministic implementation of universal quantum computation by exploiting only single-qubit gates, without requiring pre-shared large entangled states. This approach provides a solid foundation for further theoretical investigations and experimental validations aimed at exploring superposed orders of quantum gates for photonic quantum computing.

% --------------------------------------------------------------------------
% Sec. II
% --------------------------------------------------------------------------
\section{Background}
\label{sec:2}
In this section we first summarize the main challenges arising with the implementation of controlled operations in optical setups. Then, we provide the reader with a concise guide through the supermap formalism required for modeling superposed orders of quantum gates.

\begin{figure*}[t!]
	\centering
	\includegraphics[width=\linewidth]{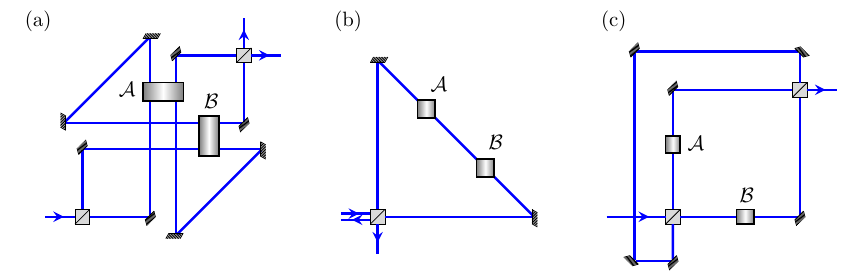}
	\caption{Schematic diagram of some of the experimentally implemented architectures of the photonic quantum switch. (a) An implementation via a Mach-Zehnder geometry, where the target qubit is encoded in polarization of the photon, while the control qubit is mapped into its path degree of freedom using the first beam splitter and coherently recombining the paths $\mathcal{A} \rightarrow \mathcal{B}$ and $\mathcal{B} \rightarrow \mathcal{A}$ at the second beam splitter~\cite{ProMoqAra-15, RubRozFei-17, Pro-19, Guo2020, RubTozMas-22}. (b) An implementation via a Sagnac geometry, where the target qubit is encoded in polarization of the photon (as in (a)), whereas a single beam splitter introduces the path degree of freedom as control and completes superposition of causal orders of $\mathcal{A}$ and $\mathcal{B}$~\cite{StrSchPet-22}. (c) An implementation via a geometry, where the target qubit is encoded in the path degree of freedom of the photon, while the role of the control qubit is played by its polarization~\cite{GosGiaKew-18, GosRomWhi-18}.}
	%\Description{Schematic diagram of some of experimentally implemented architectures of the photonic quantum switch. (a) An implementation via a Mach-Zehnder geometry, where the target qubit is encoded in polarization of the photon, while the control qubit is mapped into its path degree of freedom using the first beam splitter and coherently recombining the paths $\mathcal{A} \rightarrow \mathcal{B}$ and $\mathcal{B} \rightarrow \mathcal{A}$ at the second beam splitter~\cite{ProMoqAra-15, RubRozFei-17, Pro-19, Guo2020, RubTozMas-22}. (b) An implementation via a Sagnac geometry, where the target qubit is encoded in polarization of the photon (as in (a)), whereas a single beam splitter introduces the path degree of freedom as control and completes superposition of causal orders of $\mathcal{A}$ and $\mathcal{B}$~\cite{StrSchPet-22}. (c) An implementation via a geometry, where the target qubit is encoded in the path degree of freedom of the photon, while the role of control qubit is played by its polarization~\cite{GosGiaKew-18, GosRomWhi-18}.}
	\label{Fig:1a*}
    %\hrulefill
\end{figure*}

It is important to discuss first the physical realizability of the quantum switch. It is known that if the underlying spacetime is classical, no experiment can realise the quantum switch deterministically~\cite{Vilasini2024}. This is consistent with ~\cite{ChiDarPerVal-09} that shows the only possible circuit for a quantum switch is via a post-selection that simulates a closed time-like curve. The known experimental realizations of quantum switches overcome this limitation by implementing the parties' local operations that are delocalized in time~\cite{Oreshkov2019}. To do away with post-selection,  \emph{known experimental realizations of the quantum switch itself requires controlled unitary gates}. These controlled unitary gates correspond to arbitrary single-qubit unitary gates that require ``vacuum-extended" unitaries, i.e. in Fig. \ref{Fig:1a*}, $\mathcal{A}$ and $\mathcal{B}$ are part of $\widetilde{\mathcal{A}}=\mathcal{A}\oplus I$ and $\widetilde{\mathcal{B}}=\mathcal{B}\oplus I$~\cite{Vanrietvelde2021}. This requirement must be kept in mind in the following discussions of the implementations of two-qubit gates that use a quantum switch. Technically, because of the use of ``vacuum-extended unitaries", one may refer to experimental realizations of the quantum switch as simulations of the quantum switch. We will not dwell on that distinction in this current work.  It is sufficient to say that the operations necessary for the current proposal---a quantum switch that takes two target qubits ($Q_0$ and $Q_1$)---can be realized experimentally following extensions of the implementations shown in Fig. \ref{Fig:1a*}. Another implementation can be based on two target qubits and one control qubit on three different photons, although such realizations of the quantum switch have not been shown.

% --------------------------------------------------------------------------
\subsection{Optical Controlled Operations}
\label{sec:2.1}

It is widely recognized by both the industrial and academic communities that light represents the prominent candidate for quantum information carriers \cite{RecZeiBer-94}. In more detail, the advantages arising from photons as quantum information carriers are manifold.

Photons hardly interact with the environment, and they do not require a complex cooling infrastructure or high vacuum chambers. Additionally, they support long-range transmission with low losses through optical fiber channels or waveguides. Photons exhibit multiple degrees of freedom (DoFs) that represent a resource for quantum information encoding. While many physical realizations use polarization, photons can also exist in a superposition of time bins, frequency bins, path, and transverse modes, which can also serve as qudits \cite{WanSciLai-20}.

Unfortunately, regardless of the appealing features for quantum communications, photonic quantum technologies still represent a challenge from a quantum computing perspective. Indeed, such technologies suffer from the probabilistic nature of single-photon sources and photon-photon nonlinear interactions, which realize quantum logic gates.

However, \textbf{gate-based} photonic quantum computing can also be realized by exploiting linear optical elements \cite{KniLafMil-01}. Specifically, quantum logic operations can be obtained probabilistically through linear optical elements, ancillary photons and post-selection based on the output of single-photon detectors \cite{BaoCheZha-07, ZeuShaTil-18, LiuWeiKwe-20, LiGuQin-21, LiuWei-23, PegHelLam-24}.  In this context, particularly challenging is the realization of the logic gate $\mathtt{CNOT}$, whose scheme for optical implementation is represented in Fig.~\ref{Fig:02} and can be summarized as follows. The two input photons (control qubit $Q_0$ and target qubit $Q_1$, respectively) and two additional ancilla photons ($a_0$,$a_1$) are combined through a linear optic network of Beam Splitters (BS) as pictorially represented in Fig.~\ref{Fig:02a}. If both the ancilla qubits are detected at the output of the optical network -- specifically, both the detectors signal a single photon detection -- then the target qubit has been successfully subjected to the $\mathtt{CNOT}$ logic operation. To better understand the nondeterministic nature of the optical implementation of the gate, we consider the well-known Knill-Laflamme-Milburn (KLM) $\mathtt{CNOT}$ scheme \cite{KniLafMil-01}, represented in Fig.~\ref{Fig:02b}.
The basic unit of the scheme is a nonlinear sign-shift gate (NS) which given the input state $\ket{\mu}= a\ket{0}+b\ket{1}+c\ket{2}$ returns the output state $\ket{\mu'}= a\ket{0}+b\ket{1}-c\ket{2}$, where $\ket{0}$, $\ket{1}$ and $\ket{2}$ denote the vacuum state, single photon state and two-photon state, respectively. 
Consider the qubits encoded through spatial modes (path), the NS gate is obtained through three BS and two number-resolving detectors. The NS gate successfully performs a heralded $\pi$ phase rotation, when exactly one photon is detected on one detector and no photons are detected at the other. This event occurs with probability $1/4$. The KLM $\mathtt{CNOT}$ gate is constructed from two NS gates, hence the $\mathtt{CNOT}$ success probability is $(1/4)^2=1/16$. We represent in Fig.~\ref{Fig:02b} the KLM $\mathtt{CNOT}$ gate with two simplified NS gates, namely, the NS gates are implemented through one BS and one detector. The result is still a heralded sign shift, where the detection of one photon represents the herald event, however, the success probability is slightly decreased (from 0.25 to 0.23) \cite{OkaObrHof-11}.

For overcoming the nondeterministic nature of \textit{gate-based} schemes, the so-called \textbf{cluster-state-based} photonic quantum computing has been developed \cite{SluPry-19}. The key idea is that, in the absence of deterministic two-photon operations, an initial cluster state can be built up offline using non-deterministic interactions.  Successively, the computation progresses by manipulating the cluster state via deterministic single-qubit operations through optical elements \cite{SluPry-19}.

Our framework merges both the appealing features of \textit{gate-based} and \textit{cluster-state-based} photonic computing. Specifically, our framework overcomes the limitations of the former schemes by enabling deterministic computing. And it also overcomes the limitations exhibited by the latter, since it does not require any pre-shared multipartite-entangled state, whose generation requires offline non-deterministic interactions. 

% --------------------------------------------------------------------------
\subsection{Superposed Orders of Quantum  Operations via Quantum Switch}
\label{sec:2.2}

While the quantum circuit model is one of the most widely used paradigms of quantum computations, a lot of effort has been put into extending it to computations of higher order. Such objects as quantum combs, which can be seen as quantum circuits with open slots for arbitrary quantum gates, have allowed one to solve problems not achievable with the quantum circuit model \cite{ChiDarPer-09, ChiDarPer-09-Net, BisPer-19}. Nevertheless, quantum combs, which put the gates into a certain order on a circuit board, are not the most general computational model of higher order that can be achieved within quantum mechanics. Indeed, it allows for higher-order operations putting quantum gates into configurations -- such as the superpositions of causal orders discussed in the following -- that cannot be reduced to their well-ordered compositions \cite{ChiDarPerVal-09, OreCosBru-12, ChiDarPer-13, BisPer-19, ApaBisPer-22}. 

Superposition of causal orders of quantum operations can be realized via the quantum switch, i.e., a quantum device already implemented in numerous table-top optical and NMR experiments \cite{ProMoqAra-15, RubRozFei-17, GosGiaKew-18, Pro-19, Guo2020, GosRomWhi-18, GosRom-20, RubTozMas-22, Cao2022, Nie2022_Exp, StrSchPet-22, AntQuiWal-23, RozStrCao-24} as schematized in Fig.~\ref{Fig:1a*}. In what follows, we provide the reader with the mathematical description of the quantum switch.

Given input and output systems $\mathbf{I}$ and $\mathbf{O}$, any quantum operation transforming the former to the latter can be represented by a completely positive trace-preserving (CPTP) map $\mathcal{A}: \mathcal{L}(\mathcal{H}_\mathbf{I}) \rightarrow \mathcal{L}(\mathcal{H}_\mathbf{O})$, where $\mathcal{H}_{\mathbf{I}/\mathbf{O}}$ denotes Hilbert space of system $\mathbf{I}$ or $\mathbf{O}$, respectively, while $\mathcal{L}(\mathcal{H}_{\mathbf{I}/\mathbf{O}})$ is the set of density operators over $\mathcal{H}_{\mathbf{I}/\mathbf{O}}$.

The quantum switch is an example of a supermap that sends any $N$ quantum operations $\mathcal{A}_1[\cdot], ..., \mathcal{A}_N[\cdot]$ to a new quantum operation $\mathcal{S}(\mathcal{A}_1, ..., \mathcal{A}_N)[\cdot]$. More into details, given an ancillary quantum system $\mathbf{C}$, the quantum switch is constructed as a supermap that uses a state $\omega$ of $\mathbf{C}$ to coherently control the order in which $\mathcal{A}_1[\cdot], ..., \mathcal{A}_N[\cdot]$ act on the input system in the state $\rho$:
\begin{equation}
    \label{eq:01}
    \mathcal{S}(\mathcal{A}_1, ..., \mathcal{A}_N)[\rho \otimes \omega] = \sum_{i_1 ... i_N} K_{i_1 ... i_N} (\rho \otimes \omega) K^\dagger_{i_1 ... i_N},
\end{equation}
with
\begin{equation}
    \label{eq:02}
    K_{i_1 ... i_N} = \sum_k \mathcal{P}_k (A^{(1)}_{i_1} ... A^{(N)}_{i_{N}}) \otimes |k\rangle \langle k|    
\end{equation}
denoting the Kraus operators of the output quantum operation of the quantum switch. In \eqref{eq:02}, $\{A^{(j)}_{i_j}\}_i$ denotes the set of Kraus operators of $\mathcal{A}_j[\cdot]$, $\mathcal{P}_k$ denotes a $k$-th permutation, and $|k\rangle$ is the $k$-th basis state of system $\mathbf{C}$.

By restricting the number of controlled operations to two operations $\mathcal{A}[\cdot]$ and $\mathcal{B}[\cdot]$, the supermap in \eqref{eq:01} exhibits Kraus operators given by $K_{ij} = A_i B_j \otimes |0\rangle \langle 0| + B_j A_i \otimes |1\rangle \langle 1|$, with $\{A_i\}_i$ and $\{B_j\}_j$ being Kraus operators of $\mathcal{A}[\cdot]$ and $\mathcal{B}[\cdot]$. This new operation implemented by the quantum switch can be represented explicitly in a simple form as \cite{Chiribella2021, Simonov2022, Gao2023}:
\begin{align}
    \label{eq:03}
    \nonumber \mathcal{S}(\mathcal{A}, \mathcal{B})[\rho \otimes \omega] = & \frac{1}{4} \sum_{ij} \Bigl( \{A_i, B_j\} \rho \{A_i, B_j\}^\dagger \otimes \omega \\
    \nonumber &+ \{A_i, B_j\} \rho [A_i, B_j]^\dagger \otimes \omega Z \\
    \nonumber  &+ [A_i, B_j] \rho \{A_i, B_j\}^\dagger \otimes Z\omega \\
    &+ [A_i, B_j] \rho [A_i, B_j]^\dagger \otimes Z\omega Z \Bigr),
\end{align}
where $[\cdot, \cdot]$ and $\{\cdot, \cdot\}$ denote a commutator and an anti-commutator \cite{NieChu-11}, respectively, and $Z = |0\rangle\langle 0|-|1\rangle\langle 1|$ is the Pauli $Z$-operator.

% --------------------------------------------------------------------------
% Sec. III
% --------------------------------------------------------------------------
\section{Controlled gates via superposed orders}
\label{sec:3}

In this section, we first exploit the quantum switch to realize superposed orders of arbitrary unitary gates. Then, we exploit these preliminary results to realize controlled gates by resorting to single-qubit gates only.

% --------------------------------------------------------------------------
\subsection{Superposed orders of arbitrary gates}
\label{sec:3.1}
    
Given two arbitrary unitary gates $A$ and $B$, the execution of each of them on the input system $\mathbf{I}$ in state $\rho$ can be seen as the action of quantum operations $\mathcal{A}[\rho] = A \rho A^\dagger$ and $\mathcal{B}[\rho] = B \rho B^\dagger$. Therefore, we can realize a  superposition of causal orders between the two gates $A$ and $B$ by simplifying \eqref{eq:03} as:
\begin{align}
    \label{eq:04}
    \nonumber \mathcal{S}(\mathcal{A}, \mathcal{B})\left[\rho \otimes \omega \right] =& \frac{1}{4} \Bigl( \{A, B\} \rho \{A, B\}^\dagger \otimes \omega \\
    \nonumber &+ \{A, B\} \rho [A, B]^\dagger \otimes \omega Z \\
    \nonumber &+ [A, B] \rho \{A, B\}^\dagger \otimes Z \omega \\
    &+ [A, B] \rho [A, B]^\dagger \otimes Z \omega Z \Bigr).
\end{align}
\eqref{eq:04} can be equivalently interpreted as the execution of a new unitary gate $S(A,B)$ -- acting on the overall system composed by the input system  $\mathbf{I}$ and the ancilla $\mathbf{C}$ -- given by:
\begin{equation}
    \label{eq:05}
    S(A,B) = \frac{1}{2} \Bigl[ \{A, B\} \otimes I_\mathbf{C} + [A, B] \otimes Z_\mathbf{C} \Bigr],
\end{equation}
where $I_\mathbf{C}$ and $Z_\mathbf{C}$ denote identity and $Z$-Pauli operators, respectively, that act on $\mathbf{C}$. For taking full advantage of the indefinite causal order among the unitaries, we set system $\mathbf{C}$ in the pure state $\omega = |+\rangle \langle +|$, where $|\pm\rangle = \frac{1}{\sqrt{2}}(|0\rangle \pm |1\rangle)$, so that the input system evolves into a even superposition of the two causal orders among the unitaries.

By assuming for the sake of simplicity that the input system $\mathbf{I}$ is in a pure\footnote{The results obtained in what follows can be straightforwardly extended to arbitrary (mixed) state $\rho$ by considering the action of $S(A,B)$ on it as $S(A,B) \rho S^\dagger(A,B)$.} initial state $\rho = \ket{\psi}\bra{\psi}$, the action of the new unitary gate $S(A,B)$ is given by:
\begin{align}
    \label{eq:06}
    \nonumber S(A,B)(|\psi\rangle \otimes |+\rangle) = \frac{1}{2} \Bigl[ & \left( \{A, B\} \ket{\psi} \right) \otimes \ket{+} \\
    &+\left( [A, B]\ket{\psi} \right) \otimes \ket{-} \Bigr].
\end{align}
It is straightforward to see that tracing out the ancillary qubit $\mathbf{C}$ results in a probabilistic gate that chooses between well-ordered sequences $AB$ or $BA$ randomly with probability $1/2$. On the other hand, a measurement of the state of $\mathbf{C}$ in the basis spanned by:
\begin{align}
    \label{eq:07}
    |\mu(\theta)\rangle &= \cos\Bigl(\frac{\theta}{2}\Bigr)|0\rangle - i \sin\Bigl(\frac{\theta}{2}\Bigr) |1\rangle, \\
    \label{eq:08}
    |\mu^\perp(\theta)\rangle &= -i \sin\Bigl(\frac{\theta}{2}\Bigr)|0\rangle + \cos\Bigl(\frac{\theta}{2}\Bigr) |1\rangle, 
\end{align}
leaves the system in the following states with probability $1/2$
\begin{align}
    \label{eq:09}
    |\psi_+ (\theta)\rangle =& \Bigl[\cos\Bigl(\frac{\theta}{2}\Bigr)AB + i \sin\Bigl(\frac{\theta}{2}\Bigr) BA \Bigr]|\psi\rangle, \\
    \label{eq:10}
    |\psi_- (\theta)\rangle =& \Bigl[i\sin\Bigl(\frac{\theta}{2}\Bigr)AB + \cos\Bigl(\frac{\theta}{2}\Bigr) BA\Bigr]|\psi\rangle.
\end{align}
Accordingly, once the ancillary qubit is measured, with probability $1/2$, one of two different gates $S_+^{A,B}(\theta)$ and $S_-^{A,B}(\theta)$ is realized: 
\begin{align}
    \label{eq:11}
    S_+^{A,B}(\theta) &= \cos\Bigl(\frac{\theta}{2}\Bigr) AB + i \sin\Bigl(\frac{\theta}{2}\Bigr) BA, \\
    \label{eq:12}
    S_-^{A,B}(\theta) &= i\sin\Bigl(\frac{\theta}{2}\Bigr)AB + \cos\Bigl(\frac{\theta}{2}\Bigr)BA.
\end{align}

% --------------------------------------------------------------------------
\subsection{Superposed Orders of single-qubit gates}
\label{sec:3.2}

A pre-requisite for the realization of any universal set of quantum gates is the ability of implementing a multi-qubit gate that cannot be reduced to a single tensor product of single-qubit gates only. With the following lemma and corollary, we prove that the overall multi-qubit gate -- implemented by combining single-qubit gates via the quantum switch -- satisfies the aforementioned property for non-trivial choices of the single-qubit gates and the ancillary measurement bases.

\begin{lem}
    \label{LEM:01}
    Combining $N$ single-qubit gates $A = \bigotimes_{i=1}^N A_i$ and $N$ single-qubit gates $B = \bigotimes_{i=1}^N B_i$ via quantum switch implements one of the following two new $N$-qubit unitaries:
    \begin{align}
        \label{eq:13}
        S_+^{A,B}(\theta) &= \cos\Bigl(\frac{\theta}{2}\Bigr) \bigotimes_{i=1}^N A_i B_i + i \sin\Bigl(\frac{\theta}{2}\Bigr) \bigotimes_{i=1}^N B_i A_i , \\
        \label{eq:14}
        S_-^{A,B}(\theta) &= i\sin\Bigl(\frac{\theta}{2}\Bigr) \bigotimes_{i=1}^N A_i B_i + \cos\Bigl(\frac{\theta}{2}\Bigr) \bigotimes_{i=1}^N B_i A_i,
    \end{align}
    with the actual implemented gate depending on whether the ancillary qubit is measured as \eqref{eq:07} or \eqref{eq:08}.
    \begin{proof}
     See Appendix~\ref{app:lemma1}.
    \end{proof}
\end{lem}

We are now ready to provide the main result of this section with the following corollary.

\begin{cor}
    \label{cor}
    $S_+^{A,B}(\theta)$ and $S_-^{A,B}(\theta)$ in \eqref{eq:13} and \eqref{eq:14} cannot be reduced to a single tensor product of single-qubit gates only, unless either: i) $\theta = \pi k$ with $k \in \mathbb{Z}$, or ii) $[A_i, B_i] = 0$ for any $N-1$ gates $A_i$ and $B_i$.
    \begin{proof}
     The proof follows directly from Lemma~\ref{LEM:01}.
    \end{proof}
\end{cor}

% --------------------------------------------------------------------------
\subsection{Realization of controlled gates}
\label{sec:3.3}

\begin{figure}[t!]
    \centering
        %\begin{adjustbox}{width=0.5\columnwidth}
         %   \begin{tikzcd}
          %      &\lstick{$Q_0$} & \ctrl{2} & \qw \\
           %     && & & \\
            %    &\lstick{$Q_1$} & \gate{U} & \qw 
            %\end{tikzcd}
        %\end{adjustbox}
         \resizebox{0.4\hsize}{!}{
        \includegraphics[width=\linewidth]{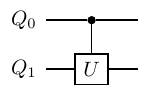}
        }
    \caption{The $\mathtt{CU}$ (controlled-$\mathtt{U}$) logic gate.}
    %\Description{The $\mathtt{CU}$ (controlled-$\mathtt{U}$) logic gate.}
    \label{fig:03}
    %\hrulefill
\end{figure}

The realization of controlled gates is key in the quantum domain. Indeed, not only entanglement is usually generated within the quantum circuit model via controlled gates (typically, using a controlled-not $\mathtt{CNOT}$), but -- even more relevant from our perspective -- there exists a class of two-qubit controlled gates any one of which is universal for quantum computation \cite{Barenco1995}.

For this, in the following we restrict our attention on two-qubit controlled gates. Accordingly, we denote the controlled gate for an arbitrary single-qubit unitary gate $\mathtt{U}$ as $\mathtt{CU}$ (controlled-$\mathtt{U}$), which is formally defined as:
\begin{equation}
    \label{eq:15}
    \mathtt{CU} = |0\rangle\langle 0| \otimes I + |1\rangle\langle 1| \otimes \mathtt{U}.
\end{equation}
Gate $\mathtt{CU}$ acts on two input qubits, with the qubit acting as control denoted as $Q_0$ and the qubit acting as target denoted with $Q_1$, as depicted in Fig.~\ref{fig:03}.

\begin{figure}[t!]
    \centering
        %\resizebox{0.8\hsize}{!}{
        \includegraphics[width=\linewidth]{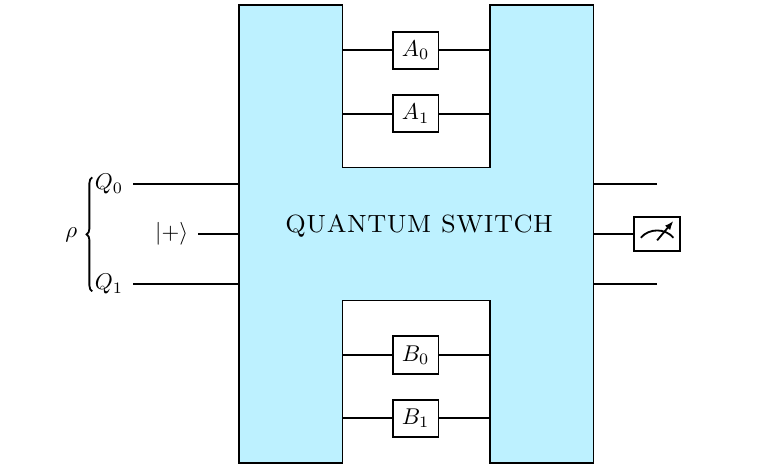}
        %}
    \caption{Representation of two-qubit controlled logic gate via quantum switch, with the switch represented as a H-shape blue box as in \cite{Oreshkov2019, KriChiSal-20, Milz2022, LugBarChi-23}. The quantum switch combines two-qubit gates $A = A_0 \otimes A_1$ and $B = B_0 \otimes B_1$ in superposed orders, with $A_0,B_0$ denoting the single-qubit unitaries acting on the first qubit $Q_0$ and $A_1,B_1$ denoting the single-qubit unitaries acting on the second qubit $Q_1$.}
    %\Description{Representation of two-qubit controlled logic gate via quantum switch, with the switch represented as a H-shape blue box as in \cite{Oreshkov2019, KriChiSal-20, Milz2022, LugBarChi-23}. The quantum switch combines two-qubit gates $A = A_0 \otimes A_1$ and $B = B_0 \otimes B_1$ in superposed orders, with $A_0,B_0$ denoting the single-qubit unitaries acting on the first qubit $Q_0$ and $A_1,B_1$ denoting the single-qubit unitaries acting on the second qubit $Q_1$.}
    \label{fig:03b}
    %\hrulefill
\end{figure}

In the following, we aim at proving that any arbitrary $\mathtt{CU}$ can be realized with the gates in \eqref{eq:13}-\eqref{eq:14}, namely, by combining single-qubit gates via the quantum switch.  Accordingly, given the two input qubits in an initial state $\rho$, the quantum switch combines two-qubit gates $A = A_0 \otimes A_1$ and $B = B_0 \otimes B_1$ in superposed orders. As represented in Fig.~\ref{fig:03b}, it is worthwhile to note that $A_0,B_0$ denote the single-qubit unitaries acting on the first qubit $Q_0$ of the input state $\rho$, whereas $A_1,B_1$ denote the single-qubit unitaries acting on the second qubit $Q_1$ of the input state $\rho$.

The following preliminary definitions are needed.

\begin{defin}
    A single-qubit gate $R_{\mathbf{n}}(\theta)$ that performs a rotation on angle $\theta$ around the $\mathbf{n}$-axis defined by the Bloch vector $\mathbf{n}=(n_X,n_Y,n_Z)$ is given by:
    \begin{equation}
        \label{eq:16}
        R_{\mathbf{n}}(\theta) = \cos\Bigl(\frac{\theta}{2}\Bigr)I - i \sin\Bigl(\frac{\theta}{2}\Bigr) (\mathbf{n} \cdot \bm{\sigma}),
    \end{equation}
    with $I$ and $\bm{\sigma} = (X, Y, Z)$ denoting the identity matrix and a vector of Pauli matrices, respectively.
\end{defin}
Accordingly, gates $R_X(\theta)$, $R_Y(\theta)$, and $R_Z(\theta)$ denote the rotation gate given in \eqref{eq:16} with respect to Bloch vectors $\mathbf{n} = (1, 0, 0)$, $\mathbf{n} = (0, 1, 0)$, and $\mathbf{n} = (0, 0, 1)$, respectively.

\begin{defin}
    A two-qubit rotation gate $R_{\mathbf{\tilde{n}}\mathbf{n}}(\theta)$ with respect to angle $\theta$ and Bloch vectors $\mathbf{\tilde{n}}$ and $\mathbf{n}$ is given by:
    \begin{equation}
        \label{eq:17}
        R_{\mathbf{\tilde{n}}\mathbf{n}}(\theta) = \cos\Bigl(\frac{\theta}{2}\Bigr)I \otimes I - i \sin\Bigl(\frac{\theta}{2}\Bigr) (\mathbf{\tilde{n}} \cdot \bm{\sigma}) \otimes (\mathbf{n} \cdot \bm{\sigma}).
    \end{equation}
\end{defin}

\begin{figure}[t!]
    \centering
        \resizebox{0.9\hsize}{!}{
        \includegraphics[width=\linewidth]{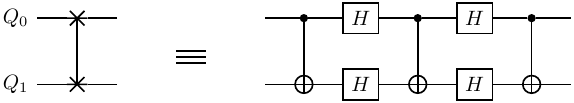}
        }
    \caption{Example of equivalent quantum circuit for a $\mathtt{SWAP}$ gate exploiting the $\mathtt{CNOT}$ logic gate implementation.}
    \label{Fig:06a}
\end{figure}

\begin{figure*}[t!]
	\centering
    \includegraphics[width=\linewidth]{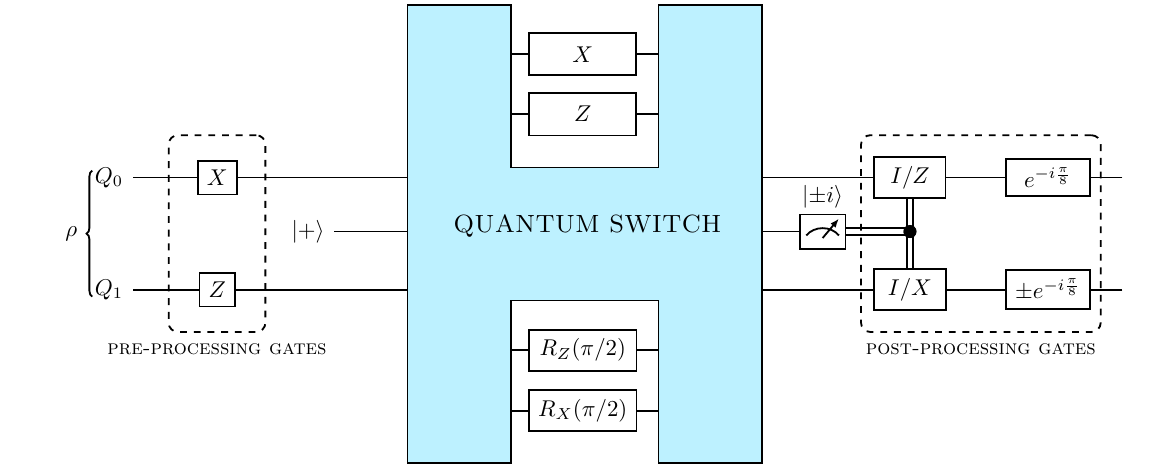}
    \caption{Abstract representation of the $\mathtt{CNOT}$ gate implemented via superposed orders of single-qubit gates, with $Q_0$ denoting the control qubit and $Q_1$ denoting the target qubit, respectively. Specifically, the order between the gates is controlled by an ancillary qubit in state $\ket{+}$ that -- after a pre-processing phase, implemented by single-qubit gates $P = X \otimes Z$ -- implements an even superposition of causal orders between single-qubit gates $A = X \otimes Z$ and $B = R_Z(\frac{\pi}{2}) \otimes R_X(\frac{\pi}{2})$. Once the ancillary qubit is measured in the basis spanned by $\ket{+i} = \frac{1}{\sqrt{2}} (\ket{0} - i\ket{1})$ and $\ket{-i} = \frac{1}{\sqrt{2}} (-i\ket{0} + \ket{1})$, qubits $Q_0$ and $Q_1$ are post-processed by $F_- = - e^{-i\frac{\pi}{4}} \left(Z \otimes X\right)$ or $F_+ = e^{-i\frac{\pi}{4}} \left(I \otimes I\right)$, depending on whether the ancilla is measured as $\ket{-i}$ or $\ket{+i}$.}
    %\Description{Abstract representation of the $\mathtt{CNOT}$ gate implemented via superposed orders of single-qubit gates, with $Q_0$ denoting the control qubit and $Q_1$ denoting the target qubit, respectively. Specifically, the order between the gates is controlled by an ancillary qubit in state $\ket{+}$ that -- after a pre-processing phase, implemented by single-qubit gates $P = X \otimes Z$ -- implements an even superposition of causal orders between single-qubit gates $A = X \otimes Z$ and $B = R_Z(\frac{\pi}{2}) \otimes R_X(\frac{\pi}{2})$. Once the ancillary qubit is measured in the basis spanned by $\ket{+i} = \frac{1}{\sqrt{2}} (\ket{0} - i\ket{1})$ and $\ket{-i} = \frac{1}{\sqrt{2}} (-i\ket{0} + \ket{1})$, qubits $Q_0$ and $Q_1$ are post-processed by $F_- = - e^{-i\frac{\pi}{4}} \left(Z \otimes X\right)$ or $F_+ = e^{-i\frac{\pi}{4}} \left(I \otimes I\right)$, depending on whether the ancilla is measured as $\ket{-i}$ or $\ket{+i}$.}
    \label{Fig:05}
    %\hrulefill
\end{figure*}

\begin{defin}
    \label{def:03}
Two-qubit gates $A$ and $B$ are locally equivalent\footnote{We note that the notion of equivalence given in Def.~\ref{def:03}, also referred to in literature as \textit{LU equivalence} \cite{Zhang2003, Musz2013, Shen2022}, restricts the allowed unitary operators to tensor product of single-qubit unitary gates only. The rationale for this constraint lies in the aim of enabling universal quantum computing via superposed orders of single qubits gates.} if they can be mapped to one another by a tensor product of single-qubit unitary gates $\{V_i\}_{i=1,2}$ and $\{\tilde{V}_i\}_{i=1,2}$:
\begin{equation}
    \label{eq:18}
    A = (V_1 \otimes V_2 ) B ( \tilde{V}_1 \otimes \tilde{V}_2 ).
\end{equation}
\end{defin}

\begin{lem}
    \label{LEM:02}
    The controlled two-qubit gate $\mathtt{CU}$ given by:
    \begin{equation}
        \label{eq:19}
        \mathtt{CU} = |0\rangle\langle 0| \otimes I + |1\rangle\langle 1| \otimes \mathtt{U},
    \end{equation}
    is locally equivalent to sequences of single-qubit gates put into superposed orders via quantum switch.
    \begin{proof}
    See Appendix~\ref{app:LEM:02}.
\end{proof}
\end{lem}

Lemma~\ref{LEM:02} shows that any two-qubit controlled gate can be implemented using \textit{single-qubit gates only} by combining them in superposed orders. By exploiting this result, the following Theorem provides a recipe for the implementation of any arbitrary $\mathtt{CU}$ gate.

\begin{theo}
    \label{theo:CU}
    The arbitrary two-qubit controlled gate $\mathtt{CU}(\alpha, \theta, \mathbf{n})$
    \begin{equation}
    \label{eq:20}
        \mathtt{CU}(\alpha, \theta, \mathbf{n}) = |0\rangle\langle 0| \otimes I + |1\rangle\langle 1| \otimes \mathtt{U}(\alpha, \theta, \mathbf{n}),
    \end{equation}
    with $\mathtt{U}(\alpha, \theta, \mathbf{n})$ defined as: 
    \begin{equation}
        \label{eq:21}
        \mathtt{U}(\alpha, \theta, \mathbf{n}) = \operatorname{exp}\Bigl[i \Bigl(\alpha I + \theta (\mathbf{n} \cdot \bm{\sigma})\Bigr)\Bigr],
    \end{equation}
    can be \textbf{deterministically} realized through:
    \begin{itemize}
        \item%[-]
        superposed orders of single-qubit gates $A(\mathbf{n}) = A_0 \otimes A_1(\mathbf{n})$ and $B(\mathbf{n}) = B_0 \otimes B_1(\mathbf{n})$ combined via a quantum switch,
        \item%[-] 
        preceded by a pre-processing phase $P(\mathbf{n}) = P_0 \otimes P_1(\mathbf{n})$, implemented by single-qubit gates $\{P_0,P_1(\mathbf{n})\}$, 
        \item%[-] 
        followed by a post-processing phase $F_\pm(\alpha, \theta, \mathbf{n})$, implemented by single-qubit gates set accordingly to the ancillary-qubit measurement results in the basis spanned by \eqref{eq:07},
    \end{itemize}
    as follows:
    \begin{equation}
        \label{eq:22}
        \mathtt{CU}(\alpha, \theta, \mathbf{n}) = F_\pm(\alpha, \theta, \mathbf{n}) S_\pm^{A(\mathbf{n}),B(\mathbf{n})}(\theta) P(\mathbf{n}).
    \end{equation}    
    where:
    \begin{align}
        \label{eq:23}
        F_\pm(\alpha, \theta, \mathbf{n}) &= e^{i\frac{\alpha}{2}} \left( R_Z\Bigl(\alpha\pm\frac{\pi}{2}\Bigr) \otimes R_\mathbf{n}\Bigl(-\theta\pm\frac{\pi}{2}\Bigr) \right), \\
        \label{eq:24}
        A(\mathbf{n}) &= A_0 \otimes A_1(\mathbf{n}) = X \otimes (\mathbf{n}^\perp \cdot \bm{\sigma}),\\
        \label{eq:25}
        B(\mathbf{n}) &= B_0 \otimes B_1(\mathbf{n}) =R_Z\Bigl(\frac{\pi}{2}\Bigr) \otimes R_{\mathbf{n}}\Bigl(\frac{\pi}{2}\Bigr), \\
        \label{eq:26}
        P(\mathbf{n}) &= P_0 \otimes P_1(\mathbf{n}) = X \otimes (\mathbf{n}^\perp \cdot \bm{\sigma}),
    \end{align}
    with $\mathbf{n}^\perp$ denoting the Bloch vector perpendicular to $\mathbf{n}$.
    \begin{proof}
        The proof follows by exploiting the result of Lemma~\ref{LEM:02}, by plugging \eqref{eq:23}-\eqref{eq:26} into the decomposition \eqref{eq:22}, and by comparing it with the decomposition
        \begin{equation}
            \label{eq:27}
            \mathtt{CU}(\alpha, \theta, \mathbf{n}) = e^{i\frac{\alpha}{2}} \Bigl(R_Z(\alpha) \otimes R_\mathbf{n}(-\theta) \Bigr) R_{Z\mathbf{n}}(\theta)
        \end{equation}
        of the $\mathtt{CU}$ gate provided in Appendix~\ref{app:LEM:02}.
    \end{proof}
\end{theo}

\begin{rem}
    It is worthwhile to note that the deterministic realization of an arbitrary controlled gate $\mathtt{CU}$ via superposed orders of single-qubit gates imposes different constraints on the gates acting on the control qubit with respect to the gates acting on the target qubit. Specifically, the single-qubit gates acting on the target -- i.e., $A_1$ and $B_1$ -- depend on the actual definition of gate $\mathtt{U}$. Conversely, the single-qubit gates acting on the control qubit -- i.e., $A_0$ and $B_0$ -- do not depend on the gate $\mathtt{U}$.
\end{rem}

Note that the pre-processing and post-processing gates specified in Theorem~\ref{theo:CU} are single-qubit, local unitaries that map the quantum-switch construction to the exact controlled operation $\mathtt{CU}$. Omitting these local operations yields output states that differ from the target only by local unitaries, i.e., they are LU-equivalent to the output obtained when the pre- and post-processing operations are included. This shows that the essential ``two-parties'' interaction of the implemented controlled operation is entirely captured by the superposed-order construction itself.

Moreover, while the above discussion focuses on controlled two-qubit unitaries, the framework naturally extends to arbitrary two-qubit gates through compilation \cite{CalAmoFer-24}. In particular, a compiler can transpile any target unitary into a circuit composed solely of single-qubit and controlled-unitary operations, each of which can be realized deterministically using the proposed superposed-order primitives, as a consequence of the universality. Consequently, non-controlled gates such as the $\mathtt{SWAP}$ can also be implemented within our scheme. Specifically, since a $\mathtt{SWAP}$ gate is equivalently realized by a quantum circuit utilizing three CNOT operations \cite{Hashim2022}, as shown in Fig. \ref{Fig:06a}, the $\mathtt{SWAP}$ can be realized through three invocations of the superposed-order primitive combined with local single-qubit gates.

Theorem~\ref{theo:CU} is the main instrument, exploited in what follows for the realization of several important examples of two-qubit controlled gates belonging to -- or constituting alone as for the Barenco gate -- universal sets.

% --------------------------------------------------------------------------
% Sec. IV
% --------------------------------------------------------------------------
\section{Universal Quantum Computation}
\label{sec:4}

In this section, we provide several examples of popular quantum gates that are used to construct universal sets, and we detail how they can be synthesized from single-qubit gates in superposed orders.

% --------------------------------------------------------------------------
\subsection{Controlled NOT}
\label{sec:4.1}

We start by considering the gate $\mathtt{CNOT}$ (controlled-$\mathtt{NOT}$), a paramount example of a gate widely used to construct more complex gates and lying at the core of fundamental quantum protocols such as entanglement generation, quantum teleportation, and telegate \cite{CalAmoFer-24,Alm-21,AlmAla-21,AlmWilSeb-24}.

Formally, $\mathtt{CNOT}$ is a controlled-$X$ gate which acts on two qubits as:
\begin{equation}
    \label{eq:28}
    \mathtt{CNOT} = |0\rangle\langle 0| \otimes I + |1\rangle\langle 1| \otimes X.
\end{equation}
Crucially, $\mathtt{CNOT}$ appears in several universal sets as a unique multi-qubit gate together with certain single-qubit gates. Indeed, any other unitary gate can be expressed as a sequence of $\mathtt{CNOT}$ gates and some single-qubit gates \cite{NieChu-11}. Hence, an efficient implementation of the $\mathtt{CNOT}$ gate is of paramount importance when it comes to universal quantum computation. To this aim, the following proposition demonstrates that the $\mathtt{CNOT}$ gate can be realized using only simple, widely-used single-qubit gates by properly placing some of them into a superposition of orders.

\begin{prop}
    \label{prop:01}
    The $\mathtt{CNOT}$ gate can be deterministically realized by combining in superposition of orders the following single-qubit gates: 
    \begin{align}
        \label{eq:29}
        A &= X \otimes Z, \\
        \label{eq:30}
        B &= R_Z\Bigl(\frac{\pi}{2}\Bigr)  \otimes R_X\Bigl(\frac{\pi}{2}\Bigr).
    \end{align}
    \begin{proof}
        The proof follows from Theorem~\ref{theo:CU}, by recognizing that the $\mathtt{CNOT}$ gate is given by the decomposition in \eqref{eq:22} for $\alpha = -\frac{\pi}{2}$, $\theta = \frac{\pi}{2}$ and $\mathbf{n} = (1,0,0)$ and choosing $\mathbf{n}^\perp = (0,0,1)$.
    \end{proof}
\end{prop}

\begin{figure}[t!]
    \centering
        \resizebox{0.85\hsize}{!}{
        \includegraphics[width=\linewidth]{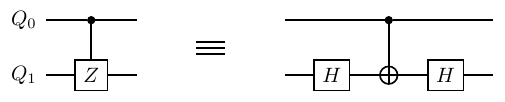}
        }
    \caption{Example of equivalent quantum circuit for a $\mathtt{CZ}$ logic gate exploiting the $\mathtt{CNOT}$ logic gate implementation.}
    \label{Fig:06}
\end{figure}

\begin{figure*}[t!]
    \centering
    \includegraphics[width=\linewidth]{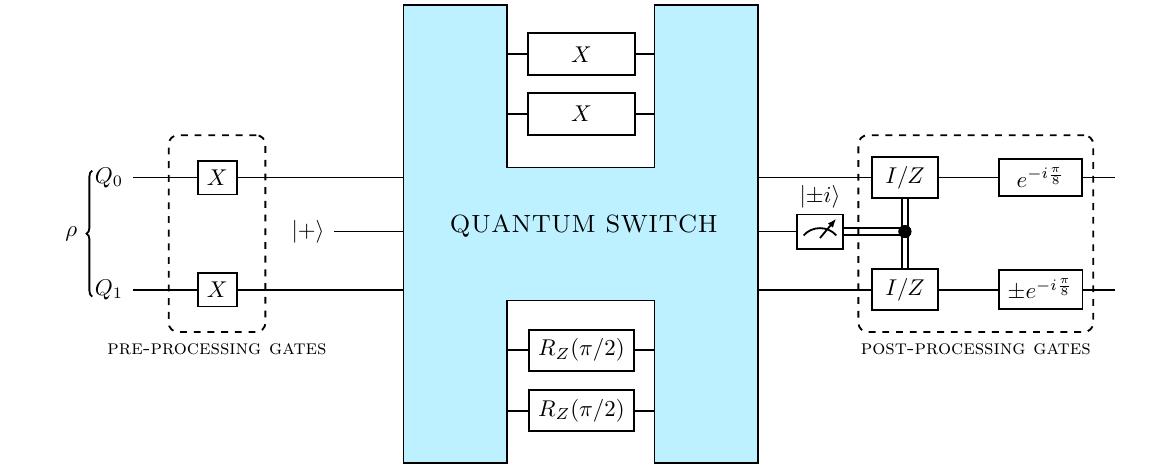}
    \caption{Abstract representation of the $\mathtt{CZ}$ gate realized via superposed orders of single-qubit gates. As in Fig.~\ref{Fig:05}, the ancillary qubit is set in $\ket{+}$ to implement an even superposition of causal orders, and the pre-processing phase is implemented by single-qubit gates $P = X \otimes X$. The single-qubit gates in superposed orders via quantum switch are $A = X \otimes X$ and $B = R_Z(\frac{\pi}{2}) \otimes R_Z(\frac{\pi}{2})$, respectively. Once the ancillary qubit is measured in the basis spanned by $\ket{+i} = \frac{1}{\sqrt{2}} (\ket{0} - i\ket{1})$ and $\ket{-i} = \frac{1}{\sqrt{2}} (-i\ket{0} + \ket{1})$, qubits $Q_0$ and $Q_1$ are post-processed by $F_{-} = - e^{-i\frac{\pi}{4}} \left(Z \otimes Z\right)$ or $F_{+} = e^{-i\frac{\pi}{4}} \left(I \otimes I\right)$, depending on whether the ancilla is measured as $\ket{-i}$ or $\ket{+i}$.}
    %\Description{Abstract representation of the $\mathtt{CZ}$ gate realized via superposed orders of single-qubit gates. As in Figure~\ref{Fig:05}, the ancillary qubit is set in $\ket{+}$ to implement an even superposition of causal orders, and the pre-processing phase is implemented by single-qubit gates $P = X \otimes X$. The single-qubit gates in superposed orders via quantum switch are $A = X \otimes X$ and $B = R_Z(\frac{\pi}{2}) \otimes R_Z(\frac{\pi}{2})$, respectively. Once the ancillary qubit is measured in the basis spanned by $\ket{+i} = \frac{1}{\sqrt{2}} (\ket{0} - i\ket{1})$ and $\ket{-i} = \frac{1}{\sqrt{2}} (-i\ket{0} + \ket{1})$, qubits $Q_0$ and $Q_1$ are post-processed by $F_{-} = - e^{-i\frac{\pi}{4}} \left(Z \otimes Z\right)$ or $F_{+} = e^{-i\frac{\pi}{4}} \left(I \otimes I\right)$, depending on whether the ancilla is measured as $\ket{-i}$ or $\ket{+i}$.}
    \label{Fig:07}
    %\hrulefill
\end{figure*}

Proposition \ref{prop:01} provides us with the realization of the $\mathtt{CNOT}$ gate via superposed orders of single-qubit gates. Specifically, as shown in Fig.~\ref{Fig:05}, we first pre-process the input state -- namely, control and target qubits $Q_0$ and $Q_1$ -- with a sequence of single-qubit gates $P = X \otimes Z$ in accordance with \eqref{eq:26}. Then, the resulting state goes into the quantum switch, which processes it according to simple, widely used single-qubit gates as in \eqref{eq:29} and \eqref{eq:30}. A measurement of the ancillary qubit $\mathbf{C}$ in the basis spanned by states \eqref{eq:07} and \eqref{eq:08} -- which we denote in this case as $\ket{+i} = \frac{1}{\sqrt{2}} (\ket{0} - i\ket{1})$ and $\ket{-i} = \frac{1}{\sqrt{2}} (-i\ket{0} + \ket{1})$ -- realizes the gates $S_\pm^{A,B}(\frac{\pi}{2})$ with probability $1/2$ in accordance with \eqref{eq:13} and \eqref{eq:14}. Finally, depending on the outcome of the measurement, we perform a post-processing by applying another sequence of single-qubit gates, namely, i) $F_-(-\pi/2,\pi/2, \mathbf{n}) = -e^{-i\frac{\pi}{4}} \left( Z \otimes X \right)$ whenever the ancillary qubit is found in the state $\ket{-i}$ or ii) the identity $F_+(-\pi/2,\pi/2, \mathbf{n})= e^{-i\frac{\pi}{4}} \left( I \otimes I \right)$ otherwise. Accordingly, the overall deterministic implementation of the $\mathtt{CNOT}$ via superposed orders can be expressed as:
\begin{equation}
    \mathtt{CNOT} = 
        \begin{cases}
            e^{-i\frac{\pi}{4}} S_+^{A,B}(\frac{\pi}{2}) (X \otimes Z ) & \text{\rm if ancilla in} \ket{+i}, \\
            -e^{-i\frac{\pi}{4}}(Z \otimes X ) S_-^{A,B}(\frac{\pi}{2}) (X \otimes Z ) & \text{\rm otherwise}.
        \end{cases} \label{eq:cnotSwitch}
\end{equation}
with, again, $A,B$ given in \eqref{eq:29} and \eqref{eq:30}.

% --------------------------------------------------------------------------
\subsection{Controlled Z}
\label{sec:4.2}

\begin{figure*}[t!]
	\centering
    \includegraphics[width=\linewidth]{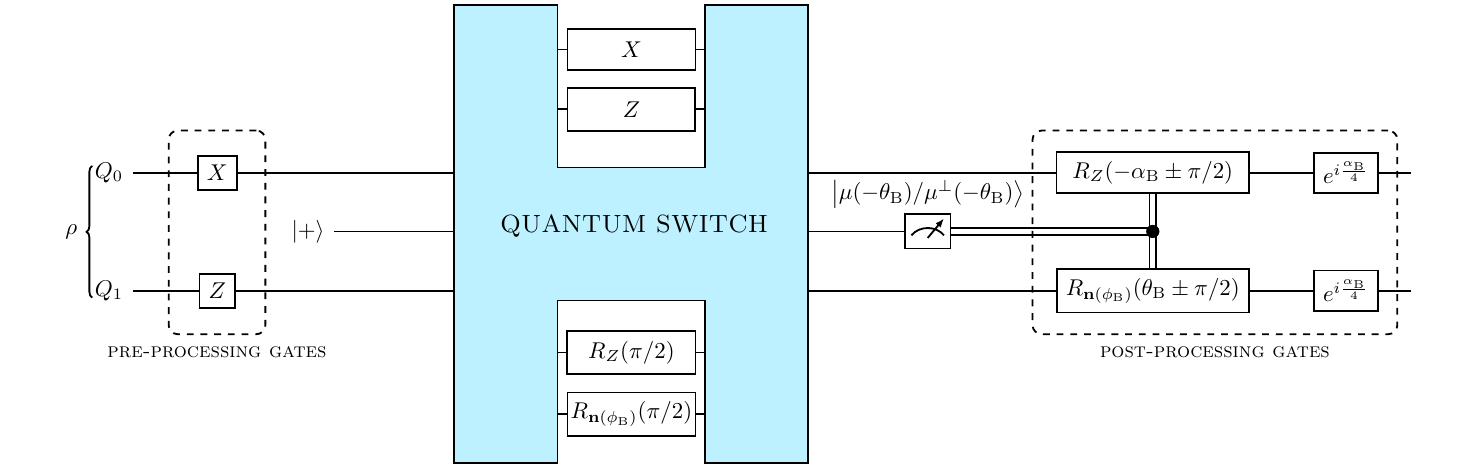}
    \caption{Abstract representation of the $\mathtt{BAR}(\alpha_{\mathtt{B}}, \phi_{\mathtt{B}}, \theta_{\mathtt{B}})$ gate realized via superposed orders of single-qubit gates. As in Fig.~\ref{Fig:05}, the ancillary qubit is set in $\omega=\ket{+}$ to implement an even superposition of causal orders, and the pre-processing phase is implemented by single-qubit gates $P = X \otimes Z$. The single-qubit gates in superposed orders via quantum switch are $A = X \otimes Z$ and $B = R_Z(\frac{\pi}{2}) \otimes R_{\mathbf{n}(\phi_{\mathtt{B}})}(\frac{\pi}{2})$, with $\mathbf{n}(\phi_{\mathtt{B}}) = (\cos(\phi_{\mathtt{B}}), \sin(\phi_{\mathtt{B}}), 0)$. Once the ancillary qubit is measured in the basis spanned by the states \eqref{eq:07} and \eqref{eq:08} with $\theta = -\theta_{\mathtt{B}}$, qubits $Q_0$ and $Q_1$ are post-processed by either $F_+ = e^{i\frac{\alpha_{\mathtt{B}}}{2}}\left(R_Z(\alpha_{\mathtt{B}} + \frac{\pi}{2}) \otimes R_{\mathbf{n}(\phi_{\mathtt{B}})}(\theta_{\mathtt{B}} + \frac{\pi}{2})\right)$ or $F_- = e^{i\frac{\alpha_{\mathtt{B}}}{2}}\left(R_Z(\alpha_{\mathtt{B}} - \frac{\pi}{2}) \otimes R_{\mathbf{n}(\phi_{\mathtt{B}})}(\theta_{\mathtt{B}} - \frac{\pi}{2})\right)$, depending on the ancilla qubit measurement result.}
    %\Description{Abstract representation of the $\mathtt{BAR}(\alpha_{\mathtt{B}}, \phi_{\mathtt{B}}, \theta_{\mathtt{B}})$ gate realized via superposed orders of single-qubit gates. As in Figure~\ref{Fig:05}, the ancillary qubit is set in $\omega=\ket{+}$ to implement an even superposition of causal orders, and the pre-processing phase is implemented by single-qubit gates $P = X \otimes Z$. The single-qubit gates in superposed orders via quantum switch are $A = X \otimes Z$ and $B = R_Z(\frac{\pi}{2}) \otimes R_{\mathbf{n}(\phi_{\mathtt{B}})}(\frac{\pi}{2})$, with $\mathbf{n}(\phi_{\mathtt{B}}) = (\cos(\phi_{\mathtt{B}}), \sin(\phi_{\mathtt{B}}), 0)$. Once the ancillary qubit is measured in the basis spanned by the states \eqref{eq:07} and \eqref{eq:08} with $\theta = -\theta_{\mathtt{B}}$, qubits $Q_0$ and $Q_1$ are post-processed by either $F_+ = e^{i\frac{\alpha_{\mathtt{B}}}{2}}\left(R_Z(\alpha_{\mathtt{B}} + \frac{\pi}{2}) \otimes R_{\mathbf{n}(\phi_{\mathtt{B}})}(\theta_{\mathtt{B}} + \frac{\pi}{2})\right)$ or $F_- = e^{i\frac{\alpha_{\mathtt{B}}}{2}}\left(R_Z(\alpha_{\mathtt{B}} - \frac{\pi}{2}) \otimes R_{\mathbf{n}(\phi_{\mathtt{B}})}(\theta_{\mathtt{B}} - \frac{\pi}{2})\right)$, depending on the ancilla qubit measurement result.}
    \label{Fig:BAR}
    %\hrulefill
\end{figure*}

While $\mathtt{CNOT}$ gate can be directly associated with a classical reversible $\mathtt{XOR}$ gate, a logic gate without classical counterpart known as $\mathtt{CZ}$ (controlled-$Z$) gate is also frequently used due to its diagonal form and can constitute a universal set together with the corresponding single-qubit gates \cite{NieChu-11}. Formally, $\mathtt{CZ}$ acts on two qubits as 
\begin{equation}
    \label{eq:cz}
    \mathtt{CZ} = 
     |0\rangle\langle 0| \otimes I + |1\rangle\langle 1| \otimes \mathtt{Z},
\end{equation}
and is locally equivalent to the $\mathtt{CNOT}$ gate via Hadamard gates, as shown in Fig.~\ref{Fig:06}:
\begin{equation}
    \label{eq:czEquiv}
    \mathtt{CZ} = (I \otimes H) \mathtt{CNOT} (I \otimes H).
\end{equation}
The following proposition demonstrates that the $\mathtt{CZ}$ gate, similarly to the $\mathtt{CNOT}$ gate, can be synthesized directly using only single-qubit gates by putting some of them in superposed orders.
\begin{prop}
    \label{prop:cz}
    The $\mathtt{CZ}$ gate can be deterministically realized by combining in superposition of orders the following single-qubit gates: 
    \begin{align}
    \label{eq:CZA}
        A &= X \otimes X, \\
        B &= R_Z\Bigl(\frac{\pi}{2}\Bigr) \otimes R_Z\Bigl(\frac{\pi}{2}\Bigr) 
        \label{eq:CZB}.
    \end{align}
    \begin{proof}
        The proof follows from Theorem~\ref{theo:CU}, by recognizing that the $\mathtt{CZ}$ gate is given by the decomposition in \eqref{eq:22} for $\alpha = -\frac{\pi}{2}$, $\theta = \frac{\pi}{2}$ and $\mathbf{n} = (0,0,1)$, and choosing $\mathbf{n}^\perp = (1,0,0)$. 
    \end{proof}
\end{prop}

As represented in Fig.~\ref{Fig:07}, Proposition \ref{prop:cz} proves that the $\mathtt{CZ}$ gate is obtained via superposed orders of single-qubit gates as follows. We first pre-process the input state with a sequence of single-qubit gates $P = X \otimes X$ in accordance with \eqref{eq:26}, then the resulted state goes into the quantum switch, which processes it according to the gates in \eqref{eq:CZA} and \eqref{eq:CZB}. A measurement of the ancillary qubit $\mathbf{C}$ in basis spanned by the states \eqref{eq:07} and \eqref{eq:08}, which we denote in this case as $\ket{+i} = \frac{1}{\sqrt{2}} (\ket{0} - i\ket{1})$ and $\ket{-i} = \frac{1}{\sqrt{2}} (-i\ket{0} + \ket{1})$, realizes the gates $S_\pm^{A,B}(\frac{\pi}{2})$ with probability $1/2$ in accordance with \eqref{eq:13} and \eqref{eq:14}. Finally, depending on the outcome of the measurement, we perform a post-processing by applying another sequence of single-qubit gates, namely, $F_-(-\pi/2,\pi/2, \mathbf{n}) = -e^{-i\frac{\pi}{4}} \left( Z \otimes Z \right)$, if the ancillary qubit is found in the state $\ket{-i}$ or $F_+(-\pi/2,\pi/2, \mathbf{n})= e^{-i\frac{\pi}{4}} \left(I \otimes I\right)$ otherwise:
\begin{equation}
    \mathtt{CZ} =
        \begin{cases}
            e^{-i\frac{\pi}{4}} S_+^{A,B}(\frac{\pi}{2}) (X \otimes X) & \text{\rm if ancilla in} \ket{+i}, \\
            -e^{-i\frac{\pi}{4}}(Z \otimes Z ) S_-^{A,B}(\frac{\pi}{2}) (X \otimes X) & \text{\rm otherwise}. 
    \end{cases} \label{eq:czSwitch}
\end{equation}

The discussed implementation of the qubit gates $\mathtt{CNOT}$ and $\mathtt{CZ}$ naturally extends to higher-dimensional systems ($d>2$). The corresponding $d$-dimensional counterparts are \cite{Got-99, DabWanSan-03, WanHuSan-20}
\begin{eqnarray}
    \mathtt{SUM}_d &=& \sum_{k=0}^{d-1} |k\rangle\langle k| \otimes X_d^m, \label{eq:dSum} \\
    \mathtt{PHASE}_d &=& \sum_{k=0}^{d-1} |k\rangle\langle k| \otimes Z_d^m, \label{eq:dPhase}
\end{eqnarray}
where
\begin{eqnarray}
    X_d &=& \sum_{k=0}^{d-1} |k\rangle\langle k+1 \, (\operatorname{mod} d)|, \\
    Z_d &=& \sum_{k=0}^{d-1} \omega^k |k\rangle\langle k|,
\end{eqnarray}
and $\omega = e^{i\frac{2\pi}{d}}$. For $d=2$, these reduce respectively to the qubit $\mathtt{CNOT}$ and $\mathtt{CZ}$ gates. While a comprehensive discussion on higher-dimensional quantum
computation lies beyond the main scope of the present work, we provide a possible implementation of \eqref{eq:dSum} and \eqref{eq:dPhase} via the superposed-order framework in Appendix~\ref{app:PHASE_SUM_d}. As future work, we will investigate suitable strategies to reduce the number of required quantum switches for realizing controlled unitaries in higher dimensions.

% --------------------------------------------------------------------------
\subsection{Barenco gate}
\label{sec:4.3}

Although $\mathtt{CNOT}$ and $\mathtt{CZ}$ gates are widely used in quantum computing, they still require additional single-qubit gates to realize an arbitrary unitary gate, i.e., to realize universal quantum computation. Differently, there exists a gate named \textit{Barenco gate} -- denoted in the following as $\mathtt{BAR}(\alpha_{\mathtt{B}}, \phi_{\mathtt{B}}, \theta_{\mathtt{B}})$ -- which alone is sufficient for universal quantum computation \cite{Barenco1995}. In other words, Barenco gate forms by itself a universal set of quantum gates. Formally, Barenco gate is a controlled rotation gate, which acts on two qubits as 
\begin{equation}
    \label{eq:bar}
    \mathtt{BAR}(\alpha_{\mathtt{B}}, \phi_{\mathtt{B}}, \theta_{\mathtt{B}}) = |0\rangle\langle 0| \otimes I + |1\rangle\langle 1| \otimes e^{i\alpha_{\mathtt{B}}} R_{\mathbf{n}(\phi_{\mathtt{B}})}(2\theta_{\mathtt{B}}), 
\end{equation}
and is parameterized by the angles $\alpha_{\mathtt{B}}, \phi_{\mathtt{B}}, \theta_{\mathtt{B}} \in [0, 2\pi]$, with $\mathbf{n}(\phi_{\mathtt{B}}) = (\cos(\phi_{\mathtt{B}}), \sin(\phi_{\mathtt{B}}), 0)$. Though its universality, practical implementation of the Barenco gate is known to be highly challenging \cite{Williams2011, Shi2018}. The following proposition demonstrates that the $\mathtt{BAR}(\alpha_{\mathtt{B}}, \phi_{\mathtt{B}}, \theta_{\mathtt{B}})$ gate can be realized for any choice of $\alpha_{\mathtt{B}}, \phi_{\mathtt{B}}, \theta_{\mathtt{B}}$ by using only single-qubit gates in a superposition of orders.

\begin{prop}
    \label{prop:bar}
    The $\mathtt{BAR}(\alpha_{\mathtt{B}}, \phi_{\mathtt{B}}, \theta_{\mathtt{B}})$ gate can be deterministically realized by combining in superposition of orders the following single-qubit gates: 
    \begin{eqnarray}\label{eq:BarA}
        A &=& X \otimes Z, \\
        B &=& R_Z\Bigl(\frac{\pi}{2}\Bigr) \otimes R_{\mathbf{n}(\phi_{\mathtt{B}})}\Bigl(\frac{\pi}{2}\Bigr), \label{eq:BarB}
    \end{eqnarray}
    where $\mathbf{n}(\phi_{\mathtt{B}}) = (\cos(\phi_{\mathtt{B}}), \sin(\phi_{\mathtt{B}}), 0)$.
    \begin{proof}
        The proof follows from Theorem~\ref{theo:CU}, by recognizing that the $\mathtt{BAR}(\alpha_{\mathtt{B}}, \phi_{\mathtt{B}}, \theta_{\mathtt{B}})$ gate is given by the decomposition in \eqref{eq:22} for $\alpha = \alpha_{\mathtt{B}}, \theta = -\theta_{\mathtt{B}}$, and $\mathbf{n} = (\cos(\phi_{\mathtt{B}}), \sin(\phi_{\mathtt{B}}), 0) \equiv \mathbf{n}(\phi_{\mathtt{B}})$ and choosing $\mathbf{n}^{\perp} = (0, 0, 1)$.
        
\end{proof}
\end{prop}
As shown in Fig.~\ref{Fig:BAR}, Proposition~\ref{prop:bar} proves that the $\mathtt{BAR}(\alpha_{\mathtt{B}}, \phi_{\mathtt{B}}, \theta_{\mathtt{B}})$ gate is obtained via superposed orders of single-qubit gates as follows. We first pre-process the input state with a sequence of single-qubit gates $P = X \otimes Z$ in accordance with \eqref{eq:26}. Then, the resulted state goes into the quantum switch, which processes it according to the gates \eqref{eq:BarA} and \eqref{eq:BarB}. A measurement of the ancillary qubit $\mathbf{C}$ in basis spanned by the states \eqref{eq:07} and \eqref{eq:08}, with $\theta = -\theta_{\mathtt{B}}$, realizes the gates $S_\pm^{A,B}(-\theta_{\mathtt{B}})$ with probability $1/2$ in accordance with \eqref{eq:13} and \eqref{eq:14}. Finally, depending on the outcome of the measurement, we perform a post-processing by applying another sequence of single-qubit gates, namely, i) $F_+(\alpha_{\mathtt{B}}, -\theta_{\mathtt{B}}, \mathbf{n}(\phi_{\mathtt{B}})) = e^{i\frac{\alpha_{\mathtt{B}}}{2}}R_Z(\alpha_{\mathtt{B}}+\frac{\pi}{2}) \otimes R_\mathbf{n(\phi_{\mathtt{B}})}(\theta_{\mathtt{B}}+\frac{\pi}{2})$, whenever the the ancillary qubit is found in the state $\ket{\mu(-\theta_{\mathtt{B}})}$ or ii) $F_-(\alpha_{\mathtt{B}}, -\theta_{\mathtt{B}}, \mathbf{n}(\phi_{\mathtt{B}})) = e^{i\frac{\alpha_{\mathtt{B}}}{2}}\left(R_Z(\alpha_{\mathtt{B}} - \frac{\pi}{2}) \otimes R_{\mathbf{n}(\phi_{\mathtt{B}})}(\theta_{\mathtt{B}} - \frac{\pi}{2})\right)$ otherwise. Accordingly, the overall deterministic implementation of the $\mathtt{BAR}(\alpha_{\mathtt{B}}, \phi_{\mathtt{B}}, \theta_{\mathtt{B}})$ gate via superposed orders can be expressed as: 
\begin{align}
    \nonumber \mathtt{BAR}(\alpha_{\mathtt{B}}, \phi_{\mathtt{B}}, \theta_{\mathtt{B}}) =& e^{i\frac{\alpha_{\mathtt{B}}}{2}} \Bigl(R_Z\Bigl(\alpha_{\mathtt{B}}\pm\frac{\pi}{2}\Bigr) \otimes R_\mathbf{n(\phi_{\mathtt{B}})}\Bigl(\theta_{\mathtt{B}}\pm\frac{\pi}{2}\Bigr) \Bigr) \\
    &\cdot S_\pm^{A,B}(-\theta_{\mathtt{B}}) (X \otimes Z).
\end{align}

\section{Conclusions}
\label{sec:5}
In this paper, we proved that quantum gates in superposed orders via the quantum switch give birth to a novel paradigm for universal quantum computation. Specifically, the quantum switch enables a framework able to implement any two-qubit controlled gate in a deterministic manner, by exploiting only single-qubit gates in superposition of causal orders. 
The result here proposed paves the way for unleashing the advantages provided by the engineering of the unconventional quantum propagation phenomena towards a computing model based on higher-order quantum operations. And we do hope that this manuscript can fuel further research -- both theoretical and experimental -- about the powerful setup enabled by superposed orders of quantum gates for photonic quantum computing.

%\begin{acks}
\begin{acknowledgments}
    This work has been funded by the European Union under Horizon Europe ERC-CoG grant QNattyNet, n.101169850. Views and opinions expressed are however those of the author(s) only and do not necessarily reflect those of the European Union or the European Research Council Executive Agency. Neither the European Union nor the granting authority can be held responsible for them. JI acknowledges PNRR MUR NQSTI-PE00000023. JR acknowledges discussions with K. Goswami. KS acknowledges: This research was funded in whole or in part by the Austrian Science Fund (FWF) 10.55776/PAT4559623. For open access purposes, the author has applied a CC BY public copyright license to any author-accepted manuscript version arising from this submission.
\end{acknowledgments}
%\end{acks}

% --------------------------------------------------------------------------
% Appendices
% --------------------------------------------------------------------------
\appendix
    
% --------------------------------------------------------------------------
\section{Proof of Lemma~\ref{LEM:01}}
\label{app:lemma1}
For $A = \bigotimes_{i=1}^N A_i$ and $B = \bigotimes_{i=1}^N B_i$, the quantum switch implements the unitary gate given in \eqref{eq:05}, i.e.:
\begin{align}
    \label{eq:40}
    \nonumber S(A,B) &= \frac{1}{2}\Biggl[ \Bigl\{ \bigotimes_{i=1}^N A_i, \, \bigotimes_{i=1}^N B_i \Bigr\} \otimes I_\mathbf{C} \\
    &+ \Bigl[ \bigotimes_{i=1}^N A_i, \, \bigotimes_{i=1}^N B_i \Bigr] \otimes Z_\mathbf{C} \Biggr].
\end{align}
Hence, when the input $N$-qubits are in a pure initial state $\rho = \ket{\psi}\bra{\psi}$ and the ancillary system $\mathbf{C}$ is in the pure state $\omega = \ket{+}\bra{+}$, the output of the unitary in \eqref{eq:40} is equal to:
\begin{align}
    \label{eq:41}
    \nonumber S(A,B)(|\psi\rangle \otimes |+\rangle) &= \frac{1}{2} \Bigl[ \Bigl\{ \bigotimes_{i=1}^N A_i, \, \bigotimes_{i=1}^N B_i \Bigr\} \ket{\psi} \otimes \ket{+} \\
    &+ \Bigl[ \bigotimes_{i=1}^N A_i, \, \bigotimes_{i=1}^N B_i \Bigr] \ket{\psi} \otimes \ket{-} \Bigr].
\end{align}
Measurement of the ancillary qubit in the basis spanned by the states \eqref{eq:07} and \eqref{eq:08} leaves the input $N$ qubits in one of the two following states:
\begin{align}
    \label{eq:42}
    |\psi_+ (\theta)\rangle &= \Bigl[\cos\Bigl(\frac{\theta}{2}\Bigr) \bigotimes_{i=1}^N A_i B_i + i \sin\Bigl(\frac{\theta}{2}\Bigr) \bigotimes_{i=1}^N B_i A_i \Bigr]|\psi\rangle, \\
    \label{eq:43}
    |\psi_- (\theta)\rangle &= \Bigl[i\sin\Bigl(\frac{\theta}{2}\Bigr) \bigotimes_{i=1}^N A_i B_i + \cos\Bigl(\frac{\theta}{2}\Bigr) \bigotimes_{i=1}^N B_i A_i \Bigr]|\psi\rangle,
\end{align}
with probability $1/2$. It is straightforward to observe that this is equivalent to the implementation of one of gates in \eqref{eq:13} and \eqref{eq:14}, each with probability $1/2$, and hence the proof follows. 

% --------------------------------------------------------------------------
\section{Proof of Lemma~\ref{LEM:02}}
\label{app:LEM:02}

An arbitrary unitary gate $\mathtt{U}$ can be represented by an Hermitian operator as $\mathtt{U} = e^{i H}$. Accordingly, applying this representation to the $\mathtt{CU}$ gate, it results that $\mathtt{CU}$ can be expressed as:
\begin{equation}
    \label{eq:44}
    \mathtt{CU} = e^{\frac{i}{2}(I - Z) \otimes H}.
\end{equation}
For a single-qubit gate $\mathtt{U}$, the corresponding Hermitian operator $H$ can be expanded into the Pauli basis as:
\begin{equation}
    \label{eq:45}
    H = \alpha I + \theta (\mathbf{n} \cdot \bm{\sigma}) ,
\end{equation}
where $\alpha, \theta \in \mathbb{R}$, $\mathbf{n}$ denotes a real-valued unit vector known as Bloch vector \cite{NieChu-11}, and $\bm{\sigma} = (X, Y, Z)$ is a vector of Pauli matrices. Plugging \eqref{eq:45} into \eqref{eq:44}, after some algebraic manipulations, the $\mathtt{CU}$ gate can be decomposed into a sequence of gates as:
\begin{equation}
    \label{app:eq:27}
    \mathtt{CU} = e^{i\frac{\alpha}{2}} \Bigl(R_Z(\alpha) \otimes R_\mathbf{n}(-\theta) \Bigr) R_{Z\mathbf{n}}(\theta),
\end{equation}
where $R_{\mathbf{n}}(\theta) = \cos\Bigl(\frac{\theta}{2}\Bigr)I - i \sin\Bigl(\frac{\theta}{2}\Bigr) (\mathbf{n} \cdot \bm{\sigma})$ is the rotation operator with respect to the Bloch vector $\mathbf{n}$ defined in \eqref{eq:16}, $R_{Z}(\theta)$ is the rotation operator with respect to $\mathbf{n} = (0,0,1)$, and $R_{Z\mathbf{n}}(\theta)$ is the two-qubit rotation gate given in \eqref{eq:17} when Bloch vector $\mathbf{\tilde{n}}$ is set as $\mathbf{\tilde{n}} = (0, 0, 1)$. Clearly, from \eqref{eq:27}, it easy to recognize that the $\mathtt{CU}$ gate is locally equivalent to the two-qubit rotation gate $R_{Z\mathbf{n}}(\theta )$ given in \eqref{eq:17}.
\\
From this, let us demonstrate that the two-qubit rotation $R_{Z\mathbf{n}}(\theta )$ is locally equivalent to:
\begin{itemize}
    \item[-] single-qubit gates $A(\mathbf{n}) = A_0 \otimes A_1(\mathbf{n})$ and $B = B_0 \otimes B_1(\mathbf{n})$ combined in superposition of orders via a quantum switch,
    \item[-] preceded by a sequences $P(\mathbf{n}) = P_0 \otimes P_1(\mathbf{n})$ of single-qubit \textit{pre-processing} gates, 
    \item[-] followed by a sequences $F(\mathbf{n}) = F_0 \otimes F_1(\mathbf{n})$ of single-qubit \textit{post-processing} gates
\end{itemize}
Accordingly, by exploiting Lemma~\ref{LEM:01}, it results that the overall gate implemented by the above-described superposition of orders is either:
\begin{align}
    \label{eq:47}
    \nonumber S_+^{A,B}(\theta) P &= \cos\Bigl(\frac{\theta}{2}\Bigr) A_0 B_0 P_0 \otimes A_1 B_1 P_1 \\
    &+ i \sin\Bigl(\frac{\theta}{2}\Bigr) B_0 A_0 P_0 \otimes B_1 A_1 P_1, \\
    \label{eq:48}
    \nonumber S_-^{A,B}(\theta) P &= i \sin\Bigl(\frac{\theta}{2}\Bigr) A_0 B_0 P_0 \otimes A_1 B_1 P_1 \\
    &+ \cos\Bigl(\frac{\theta}{2}\Bigr) B_0 A_0 P_0 \otimes B_1 A_1 P_1,
\end{align}
depending on the measurement of the ancillary qubit $\mathbf{C}$ in the basis spanned by vectors \eqref{eq:07} and \eqref{eq:08}, where the parameter $\theta$ matches the corresponding parameter in decomposition \eqref{eq:45}.
\\
By setting the single-qubit gates as:
\begin{align}
    \label{eq:49}
    P(\mathbf{n}) &= P_0 \otimes P_1(\mathbf{n}) = X \otimes (\mathbf{n}^\perp \cdot \bm{\sigma}), \\
    \label{eq:50}
    A(\mathbf{n}) &= A_0 \otimes A_1(\mathbf{n}) = X \otimes (\mathbf{n}^\perp \cdot \bm{\sigma}), \\
    \label{eq:51}
    B(\mathbf{n}) &= B_0 \otimes B_1(\mathbf{n}) = R_Z\Bigl(\frac{\pi}{2}\Bigr) \otimes R_{\mathbf{n}}\Bigl(\frac{\pi}{2}\Bigr),
\end{align} 
where $\mathbf{n}^\perp$ denotes the Bloch vector perpendicular to $\mathbf{n}$ (i.e., $\mathbf{n}^\perp \cdot \mathbf{n} = 0$), the operations arising from the causal order $AB$ read as:
\begin{align}
    \label{eq:52}
    A_0 B_0 P_0 &= XR_Z\Bigl(\frac{\pi}{2}\Bigr)X = R_Z\Bigl(-\frac{\pi}{2}\Bigr), \\
    \label{eq:53}
    A_1(\mathbf{n}) B_1(\mathbf{n}) P_1(\mathbf{n}) &= (\mathbf{n}^\perp \cdot \bm{\sigma}) R_\mathbf{n}\Bigl(\frac{\pi}{2}\Bigr) (\mathbf{n}^\perp \cdot \bm{\sigma}) \nonumber \\
    &= R_\mathbf{n}\Bigl(-\frac{\pi}{2}\Bigr),
\end{align}
while the ones arising from the causal order $BA$ are:
\begin{align}
    \label{eq:54}
    B_0 A_0 P_0 &= R_Z\Bigl(\frac{\pi}{2}\Bigr)XX = R_Z\Bigl(\frac{\pi}{2}\Bigr), \\
    \label{eq:55}
    B_1(\mathbf{n}) A_1(\mathbf{n}) P_1(\mathbf{n}) &= R_\mathbf{n}\Bigl(\frac{\pi}{2}\Bigr)(\mathbf{n}^\perp \cdot \bm{\sigma})(\mathbf{n}^\perp \cdot \bm{\sigma}) \nonumber \\
    &= R_\mathbf{n}\Bigl(\frac{\pi}{2}\Bigr).
\end{align}
By substituting the above equations in  \eqref{eq:47} and \eqref{eq:48}, it results:
\begin{align}
    \label{eq:56}
    \nonumber S_+^{A,B}(\theta) P(\mathbf{n}) &= \cos\Bigl(\frac{\theta}{2}\Bigr) R_Z\Bigl(-\frac{\pi}{2}\Bigr) \otimes R_\mathbf{n}\Bigl(-\frac{\pi}{2}\Bigr) \\
    &+ i \sin\Bigl(\frac{\theta}{2}\Bigr) R_Z\Bigl(\frac{\pi}{2}\Bigr) \otimes R_\mathbf{n}\Bigl(\frac{\pi}{2}\Bigr), \\
    \label{eq:57}
    \nonumber S_-^{A,B}(\theta) P(\mathbf{n}) &= i\sin\Bigl(\frac{\theta}{2}\Bigr) R_Z\Bigl(-\frac{\pi}{2}\Bigr) \otimes R_\mathbf{n}\Bigl(-\frac{\pi}{2}\Bigr) \\
    &+ \cos\Bigl(\frac{\theta}{2}\Bigr) R_Z\Bigl(\frac{\pi}{2}\Bigr) \otimes R_\mathbf{n}\Bigl(\frac{\pi}{2}\Bigr).
\end{align}
Finally, by setting the single-qubit post-processing gates $F(\mathbf{n}) = F_0 \otimes F_1(\mathbf{n})$ equal either to:
\begin{equation}
    \label{eq:58}
    F_\pm(\mathbf{n}) = F_{\pm, 0} \otimes F_{\pm, 1}(\mathbf{n}) = R_Z\Bigl(\pm \frac{\pi}{2}\Bigr) \otimes R_\mathbf{n}\Bigl(\pm \frac{\pi}{2}\Bigr),
\end{equation}
depending on whether the ancillary qubit measurement result is  $|\mu(\theta)\rangle$ or $|\mu^\perp(\theta)\rangle$, the following two-qubit rotation is implemented:
\begin{equation}
    \label{eq:59}
    F_\pm(\mathbf{n}) S_\pm^{A,B}(\theta) P(\mathbf{n}) = R_{Z\mathbf{n}}(\theta).
\end{equation}
In \eqref{eq:59} we exploited the equality $R_Z(\pm\pi) \otimes R_\mathbf{n}(\pm\pi) = - Z \otimes (\mathbf{n} \cdot \bm{\sigma})$. As both $P(\mathbf{n})$ and $F_\pm(\mathbf{n})$ are tensor product of single-qubit case, it results that $S_\pm^{A,B}(\theta)$ -- hence -- is locally (single-qubit) equivalent to the two-qubit rotation gate $R_{Z\mathbf{n}}(\theta )$ via \eqref{eq:59}, which in turn is locally equivalent to the $\mathtt{CU}$ gate via \eqref{eq:27}. Hence, the proof follows. 

\section{Realization of $\mathtt{SUM}_d$ and $\mathtt{PHASE}_d$ gates}
\label{app:PHASE_SUM_d}

The gates $\mathtt{SUM}_d$ and $\mathtt{PHASE}_d$ are locally equivalent: \cite{DabWanSan-03}
\begin{equation}
    \mathtt{SUM}_d = (I \otimes \mathcal{F}_d^\dagger) \mathtt{PHASE}_d (I \otimes \mathcal{F}_d),
    \label{eq:C1}
\end{equation}
where the discrete Fourier transform is defined by
\begin{equation}
    \mathcal{F}_d = \frac{1}{\sqrt{d}}\sum_{k,k' = 0}^{d-1} 
    \omega^{kk'}|k\rangle\langle k'|, \qquad
    \omega = e^{2\pi i / d}.
    \label{eq:C2}
\end{equation}
In particular, for $d = 2$, \eqref{eq:C1} reduces to the local equivalence \eqref{eq:czEquiv} between $\mathtt{CNOT}$ and $\mathtt{CZ}$. We thus focus on the implementation of $\mathtt{PHASE}_d$, which, according to \eqref{eq:dPhase}, acts as
\begin{equation}
    \mathtt{PHASE}_d (|m\rangle \otimes |n\rangle) = \omega^{mn} |m\rangle \otimes |n\rangle.
    \label{eq:C3}
\end{equation}
Let the label of each computational basis state be expressed in binary form as
\begin{equation}
    m=\sum_{b=0}^{s-1} m_b2^b, \qquad 
    n=\sum_{c=0}^{s-1} n_c2^c,
    \label{eq:C4}
\end{equation}
where $s=\lceil\log_2 d\rceil$ is the number of binary digits required to encode $d$ distinct states, and $m_b, n_c \in \{0,1\}$ denote the $b$'th and $c$'th binary digits of $m$ and $n$, respectively. Equivalently, each label can be represented as the binary string $(m_{s-1} \ldots m_1 m_0) \equiv m$, where $m_b$ specifies the value ($0$ or $1$) of the $b$'th binary component. Using the binary expansions in \eqref{eq:C4}, the phase factor appearing in $\mathtt{PHASE}_d$ can then be expressed as
\begin{eqnarray}
    \nonumber \omega^{mn} &=& e^{i\frac{2\pi}{d}\sum_{b,c=0}^{s-1}2^{b+c} m_b n_c} \\
    &=& \prod_{b,c=0}^{s-1} e^{\,i\phi_{b,c} m_b n_c},
    \label{eq:C5}
\end{eqnarray}
where $\phi_{b,c} = \frac{2^{b+c+1}\pi}{d}$. Therefore, each bit pair $(b,c)$ contributes an independent phase whenever $m_b=n_c=1$. To promote this expression to operator form, we define projectors onto the subspaces where the $b$'th bit equals $1$:
\begin{equation}
    \Pi_b = \sum_{\{m: m_b=1\}}|m\rangle\langle m|.
    \label{eq:C6}
\end{equation}
For a fixed $b$, they act as identity on basis states 
with $m_b=1$ and annihilates those with $m_b=0$. In turn, $\Pi_b\otimes\Pi_c$ projects onto all states $|m\rangle \otimes |n\rangle$ for which $m_b=n_c=1$, contributing a phase $e^{i\phi_{b,c}}$ as in \eqref{eq:C5}. Therefore, $\mathtt{PHASE}_d$ can be decomposed as
\begin{equation}
    \mathtt{PHASE}_d = \prod_{\substack{b,c=0 \\ 2^{b+c} < d}}^{s-1} e^{i \phi_{b,c} \Pi_b \otimes \Pi_c},
    \label{eq:C7}
\end{equation}
where the condition $2^{b+c} < d$ reflects that, for dimensions
$d=2^s$ (i.e., when the qudit can be viewed as a register of $s$ qubits), the corresponding phases $\phi_{b,c} = 2^{b+c-s+1}\pi$ lead to $e^{i\phi_{b,c}\Pi_b\otimes\Pi_c} = I$ whenever $b+c \geq s$. Each exponential in \eqref{eq:C7} acts nontrivially only on those two-qudit basis states whose binary digits $(m_b,n_c)$ are both equal to one. Using $\Pi_b = \frac{1}{2}(I - \tilde{Z}_b)$, with
\begin{equation}
    \tilde{Z}_b = \sum_{m=0}^{d-1} (-1)^{m_b} |m\rangle\langle m|,
    \label{eq:C8}
\end{equation}
which acts as a Pauli-$Z$ operator on each pair of states that differ only by the $b$'th binary digit, every term in \eqref{eq:C7} can be rewritten as
\begin{equation}
    e^{i \phi_{b,c} \Pi_b \otimes \Pi_c} = e^{i\frac{\phi_{b,c}}{4}} \Bigl(e^{-i \frac{\phi_{b,c}}{4} \tilde{Z}_b} \otimes e^{-i \frac{\phi_{b,c}}{4} \tilde{Z}_c} \Bigr) e^{i \frac{\phi_{b,c}}{4} \tilde{Z}_b \otimes \tilde{Z}_c}.
    \label{eq:C9}
\end{equation}
Each exponential factor $e^{i\frac{\phi_{b,c}}{4}\tilde{Z}_b\otimes\tilde{Z}_c}$ can be implemented using a single quantum switch. We choose the single-qudit gates as
\begin{alignat}{3}
    \label{eq:C10}
    P^{(b,c)} & = P_{0}^{(b,c)} \otimes P_{1}^{(b,c)} && = \tilde{X}_b \otimes \tilde{X}_c, \\
    \label{eq:C11}
    A^{(b,c)} & = A_{0}^{(b,c)} \otimes A_{1}^{(b,c)} && = \tilde{X}_b \otimes \tilde{X}_c, \\
    \label{eq:C12}
    B^{(b,c)} & = B_{0}^{(b,c)} \otimes B_{1}^{(b,c)} && = e^{i\frac{\pi}{4}\tilde{Z}_b} \otimes e^{i\frac{\pi}{4}\tilde{Z}_c},
\end{alignat}
where the operator
\begin{eqnarray}
    \nonumber \tilde{X}_b &=& \sum_{\substack{m=0\\ m\oplus 2^b<d}}
    \bigl(|m\rangle\langle m\oplus 2^b|+|m\oplus 2^b\rangle\langle m|\bigr) \\
    &+& \sum_{\substack{m:\,m\oplus 2^b\ge d}}|m\rangle\langle m|,
    \label{eq:C13}
\end{eqnarray}
acts as a generalized bit flip on the $b$'th binary digit of $m$, and the $\oplus$ denotes bitwise addition modulo two acting on the binary representation of $m$,
\begin{equation}
    m\oplus 2^b \equiv (m_{s-1}\ldots m_{b+1}\,\overline{m_b}\,m_{b-1}\ldots m_0), \label{eq:C14}
\end{equation}
where $\overline{m_b}=1-m_b$ indicates that the $b$'th binary digit is flipped. Hence, $\tilde{X}_b$ acts as a Pauli-$X$ operator on each pair of states that differ only by the $b$'th binary digit, so that
\begin{equation}
\tilde{X}_b \tilde{Z}_b = -\tilde{Z}_b \tilde{X}_b, \qquad \tilde{X}_b^2=I, \label{eq:C15}
\end{equation}
so that
\begin{align}
    A^{(b,c)}B^{(b,c)}P^{(b,c)} &= e^{-i\frac{\pi}{4}\tilde{Z}_b}\otimes e^{-i\frac{\pi}{4}\tilde{Z}_c}, 
    \label{eq:C16}\\
    B^{(b,c)}A^{(b,c)}P^{(b,c)} &= e^{i\frac{\pi}{4}\tilde{Z}_b}\otimes e^{i\frac{\pi}{4}\tilde{Z}_c}.
    \label{eq:C17}
\end{align}
Applying Lemma~\ref{LEM:01}, the outputs of the switch are
\begin{eqnarray}
    \nonumber S_+^{A^{(b,c)},\,B^{(b,c)}}(\theta)P_{bc} &=& \cos\Bigl(\frac{\theta}{2}\Bigr)
    e^{-i\frac{\pi}{4}\tilde{Z}_b} \otimes e^{-i\frac{\pi}{4}\tilde{Z}_c} \\
    &+& i\sin\Bigl(\frac{\theta}{2}\Bigr) e^{i\frac{\pi}{4}\tilde{Z}_b} \otimes e^{i\frac{\pi}{4}\tilde{Z}_c}, \label{eq:C18} \\
    \nonumber S_-^{A^{(b,c)},\,B^{(b,c)}}(\theta)P_{bc} &=& i\sin\Bigl(\frac{\theta}{2}\Bigr) e^{-i\frac{\pi}{4}\tilde{Z}_b} \otimes e^{-i\frac{\pi}{4}\tilde{Z}_c} \\
    &+& \cos\Bigl(\frac{\theta}{2}\Bigr) e^{i\frac{\pi}{4}\tilde{Z}_b} \otimes e^{i\frac{\pi}{4}\tilde{Z}_c}. \label{eq:C19}
\end{eqnarray}
Setting the post-processing gates $F_{\pm}^{(b,c)} = e^{\pm i\frac{\pi}{4} \tilde{Z}_b} \otimes e^{\pm i\frac{\pi}{4} \tilde{Z}_c}$, we obtain
\begin{eqnarray}
    \nonumber F_{\pm}^{(b,c)} S_{\pm}^{A^{(b,c)},\,B^{(b,c)}}(\theta) P^{(b,c)} &=& \cos\Bigl(\frac{\theta}{2}\Bigr) I \otimes I \\
    \nonumber &\qquad& + i \sin\Bigl(\frac{\theta}{2}\Bigr) \tilde{Z}_b \otimes \tilde{Z}_c \\
    &=& e^{i\frac{\theta}{2} \tilde{Z}_b \otimes \tilde{Z}_c}. \label{eq:C20}
\end{eqnarray}
Comparing this with \eqref{eq:C9}, we conclude that each factor in the decomposition can be implemented by measuring the ancillary system $\mathbf{C}$ in the basis spanned by \eqref{eq:07} and \eqref{eq:08} with $\theta = \frac{\phi_{b,c}}{2}$, so that
\begin{eqnarray}
    \nonumber e^{i \phi_{b,c} \Pi_b \otimes \Pi_c} &=& e^{i\frac{\phi_{b,c}}{4}} \Bigl( e^{-i \frac{\phi_{b,c} \mp \pi}{4} \tilde{Z}_b} \otimes e^{-i \frac{\phi_{b,c} \mp \pi}{4}\tilde{Z}_c} \Bigr) \\
    &\times& S_{\pm}^{A^{(b,c)},\,B^{(b,c)}}\Bigl(\frac{\phi_{b,c}}{2}\Bigr) P^{(b,c)}. \label{eq:C21}
\end{eqnarray}
Therefore, the overall deterministic implementation of the gates $\mathtt{SUM}_d$ and $\mathtt{PHASE}_d$ via superposed orders can be expressed as
\begin{eqnarray}
    \nonumber \mathtt{SUM}_d &=& \prod_{\substack{b,c=0 \\ 2^{b+c} < d}}^{s-1} e^{i\frac{\phi_{b,c}}{4}} \Bigl( e^{-i \frac{\phi_{b,c} \mp \pi}{4} \tilde{Z}_b} \otimes \mathcal{F}_d^\dagger e^{-i \frac{\phi_{b,c} \mp \pi}{4}\tilde{Z}_c} \Bigr) \\
    &\times& S_{\pm}^{A^{(b,c)},\,B^{(b,c)}}\Bigl(\frac{\phi_{b,c}}{2}\Bigr) (\tilde{X}_b \otimes \tilde{X}_c \mathcal{F}_d),
    \label{eq:C22} \\
    \nonumber \mathtt{PHASE}_d &=& \prod_{\substack{b,c=0 \\ 2^{b+c} < d}}^{s-1} e^{i\frac{\phi_{b,c}}{4}} \Bigl( e^{-i \frac{\phi_{b,c} \mp \pi}{4} \tilde{Z}_b} \otimes e^{-i \frac{\phi_{b,c} \mp \pi}{4}\tilde{Z}_c} \Bigr) \\
    &\times& S_{\pm}^{A^{(b,c)},\,B^{(b,c)}}\Bigl(\frac{\phi_{b,c}}{2}\Bigr) (\tilde{X}_b \otimes \tilde{X}_c),
    \label{eq:C23}
\end{eqnarray}
where each $S_{\pm}^{A^{(b,c)},\,B^{(b,c)}}\Bigl(\frac{\phi_{b,c}}{2}\Bigr)$ is implemented by one use of the quantum switch, so that the total number of required uses is
\begin{equation}
    N = \begin{cases} \frac{s(s+1)}{2} & \text{\rm if } d = 2^s, \\
            s^2 & \text{\rm otherwise}.\end{cases} \label{eq:C24}
\end{equation}
For $d=2$, one has $s=1$, so that $b,c=0$ and $\phi_{b,c}=\pi$. In this case, the operators $\tilde{Z}_{b/c}$ and $\tilde{X}_{b/c}$ reduce to the standard Pauli operators $Z$ and $X$, while the discrete Fourier transform becomes $\mathcal{F}_2 = H$. Consequently, \eqref{eq:C22} and \eqref{eq:C23} recover the qubit implementations \eqref{eq:cnotSwitch} and \eqref{eq:czSwitch}, respectively.

\color{black}

%\bibliographystyle{ACM-Reference-Format}
%\bibliographystyle{apsrev4-1}
\bibliography{main.bib}


\end{document}